\documentclass[aps,prx,twocolumn,superscriptaddress,showpacs,longbibliography]{revtex4-1}

\usepackage{amsmath,amssymb}
\usepackage{bm}
\usepackage{braket}
\usepackage{multirow}

\usepackage{graphicx}
\usepackage{color}
\usepackage[colorlinks=true, linkcolor=magenta,citecolor=blue,urlcolor=magenta]{hyperref}
\DeclareGraphicsExtensions{.eps}	

\newcommand{\Tr}[1]{\text{Tr}[#1]}
\newcommand{\TrB}[1]{\text{Tr}_{\text{B}} [#1]}

\def\oH{\hat{H}}
\def\vH{\check{H}}

\def\oS{\hat{S}}
\def\dS{\hat{S}^{\dagger}}
\def\vS{\check{S}}
\def\dvS{\check{S}^{\dagger}}

\def\oB{\hat{B}}

\def\vB{\check{B}}

\def\ou{\hat{u}}
\def\du{\hat{u}^{\dagger}}

\def\oO{\hat{O}}

\def\vO{\check{O}}

\def\oRho{\hat{\rho}}
\def\vRho{\check{\rho}}

\def\oPi{\hat{\varPi}}

\def\oSgm{\hat{\sigma}}

\def\ob{\hat{b}}
\def\db{\hat{b}^{\dagger}}

\def\oL{\hat{\mathcal{L}}}
\def\vL{\check{\mathcal{L}}}

\def\oD{\hat{\mathcal{D}}}
\def\vD{\check{\mathcal{D}}}

\def\T{\mathcal{T}}

\def\kB{k_{\text{B}}}

\def\R{\text{R}}
\def\S{\text{S}}
\def\B{\text{B}}
\def\A{\text{A}}
\def\I{\text{I}}

\def\X{\text{X}}
\def\Y{\text{Y}}

\def\E{\text{E}}
\def\G{\text{G}}

\def\c{\text{c}}
\def\peak{\text{peak}}
\def\ad{\text{ad}}
\def\LZ{\text{LZ}}
\def\Hc{\mathrm{H.c.}}

\def\ii{\mathrm{i}}
\def\dd{\mathrm{d}}

\def\M{\text{M}}
\def\oX{\hat{X}}

\def\oL{\hat{\mathcal{L}}}
\def\dL{\hat{\mathcal{L}}^{\dagger}}

\def\vL{\check{\mathcal{L}}}

\def\vG{\check{\mathcal{G}}}
\def\vK{\check{\mathcal{K}}}
\def\vI{\check{\mathcal{I}}}
\def\vW{\check{\mathcal{W}}}

\def\oU{\hat{\mathcal{U}}}
\def\dU{\hat{\mathcal{U}}^{\dagger}}
\def\vU{\check{\mathcal{U}}}

\def\oP{\mathcal{P}}
\def\oQ{\mathcal{Q}}

\def\oT{\mathcal{T}}

\def\SLP{\hat{\mathcal{S}}}
\def\oC{\hat{\mathcal{C}}}

\begin{document}
\title{Markovian Quantum Master Equation beyond Adiabatic Regime}

\author{Makoto Yamaguchi}
\altaffiliation{E-mail: makoto.yamaguchi@riken.jp}
\affiliation{Center for Emergent Matter Science, RIKEN, Wakoshi, Saitama 351-0198, Japan}
\author{Tatsuro Yuge}
\affiliation{Department of Physics, Shizuoka University, Shizuoka 422-8529, Japan}
\author{Tetsuo Ogawa}
\affiliation{Department of Physics, Osaka University, 1-1 Machikaneyama, Toyonaka, Osaka 560-0043, Japan}
\date{\today}

\begin{abstract}
By introducing a temporal change timescale $\tau_{\A}(t)$ for the time-dependent system Hamiltonian, a general formulation of the Markovian quantum master equation is given to go well beyond the adiabatic regime.
In appropriate situations, the framework is well justified even if $\tau_{\A}(t)$ is faster than the decay timescale of the bath correlation function.
An application to the dissipative Landau-Zener model demonstrates this general result.
The findings are applicable to a wide range of fields, providing a basis for quantum control beyond the adiabatic regime.
\end{abstract}

\maketitle

\section{Introduction}\label{sec:Intro}
\begin{figure}[!tb] 
\centering
\includegraphics[width=.49\textwidth]{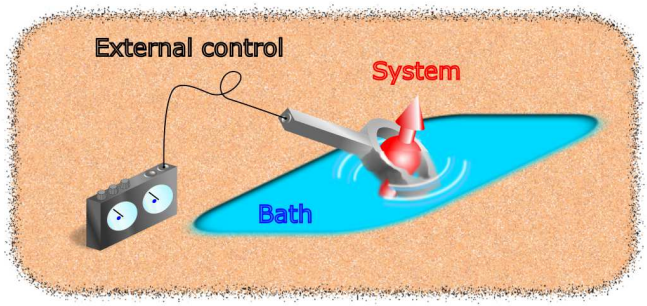}
\caption{(Color online)
Schematic illustration of an open quantum system under the external control.
The system, depicted by a (collective) spin, is in a bath and controlled by external field.
The external field is in general time-dependent in order to make use of the quantum system as a physical resource, e.g. for the AQC and the QA.
}
\label{fig:Schematic}
\end{figure}

The Markovian quantum master equation (QME)~\cite{Breuer02, Rivas12} provides a key paradigm for the study of nonequilibrium statistical physics.
The structure of this framework is transparent, the generality of its derivation is sufficient, and the approximations applied are well defined, especially when the system Hamiltonian is time-independent.
In the last decades, therefore, the Markovian QME has allowed a broad range of applications for the study of open quantum systems that have {\em time-independent} system Hamiltonians~\cite{Carmichael93, Scully97, Wichterich07, Esposito09, Werlang15,Yuge14,Jin14}.
In recent years, however, the quantum dynamics driven by {\em time-dependent} external field (see also Fig.~\ref{fig:Schematic}) has been an area of growing importance in various contexts, such as adiabatic quantum computation (AQC)~\cite{Farhi01, Amin09-1, Amin09-2}, quantum annealing (QA)~\cite{Kadowaki98, Dickson13, Pudenz14, Amin08, Childs01, Vega10, Albash12}, quantum heat engines~\cite{Alicki79,Kosloff02,Uzdin15}, Bose-Einstein condensates in optical lattices~\cite{Chen11,Bason12}, and semiconductor quantum dots~\cite{Petta10,Petersson10,Cao13} because the external driving is essential for the use of the quantum systems as nontrivial physical resources.
Nevertheless, no general way of rigorously constructing the Markovian QME is currently known for the time-dependent system Hamiltonian, $\oH_{\S}(t)$~\cite{Rivas12}.
One therefore often comes across a fundamental problem of describing the open quantum dynamics simultaneously subject to the external driving.

One possible way to circumvent this difficulty is the Floquet formalism if the system Hamiltonian is driven periodically in time~\cite{Kohler97, Grifoni98, Kamleitner11, Forster15}.
However, this approach is obviously inadequate in situations with no periodicity.
Another natural approach is then to assume the slow temporal change of $\oH_{\S}(t)$.
However, the following two questions immediately arise in turn: how can we quantify the temporal change timescale of $\oH_{\S}(t)$, which we denote by $\tau_{\A}(t)$, and what is the other timescale to be compared with $\tau_{\A}(t)$ when we say slow?
In relation to these questions, various different statements can be found in literature.
In particular, an important assumption employed by many authors is the adiabatic regime that satisfies the ordinary adiabatic theorem~\cite{Born28,Kato50}, in which any non-adiabatic transitions between instantaneous eigenstates are suppressed~\cite{Davies78, Childs01, Vega10, Cai10, Albash12, Kamleitner13, Pekola10,Vega15}.
However, under this assumption, any non-adiabatic effect cannot be discussed despite the recent experimental demand;
in the context of the QA, for example, it is essential to consider the non-adiabatic transitions with dissipation in practice~\cite{Dickson13}. 
Alternatively, another assumption sometimes encountered is the slow temporal change of $\oH_{\S}(t)$ compared to the decay timescale of the bath correlation function, $\tau_{\B}$.
However, due to the lack of the general definition of $\tau_{\A}(t)$, there is no consensus for this condition even if simple systems, e.g.~the dissipative Landau-Zener (DLZ) models, are considered~\cite{Whitney11, Nalbach14, Javanbakht15}.
The conditions for the Markovian QME are thus subject to long-standing debate when the system Hamiltonian depends on time, originally started from Davies and Spohn~\cite{Davies78}.

In this article, our purpose is two-fold.
First, under the weak-coupling approximation (WCA), we settle this long-standing debate by studying the conditions to justify the Markovian QME with $\oH_{\S}(t)$ in a general manner.
We introduce an explicit definition of $\tau_{\A}(t)$ and approximations required to obtain the QME under the WCA.
As a result, we shall see that the Markovian QME is naturally derived {\it without the adiabatic theorem}.
Our route of the formulation successfully removes the ambiguities of the relevant conditions.
Second, in a broad range of situations, we further find that there is no need to even assume the slow temporal change of $\oH_{\S}(t)$ compared to $\tau_{\B}$.
This is in contrast to the common belief that the Markovian description breaks down when the system Hamiltonian changes more rapidly than $\tau_{\B}$~\cite{Davies78, Vega10, Cai10, Albash12, Kamleitner13, Pekola10,Vega15,Whitney11, Nalbach14, Javanbakht15}.
In consequence, the non-adiabatic regime becomes definitely accessible by the Markovian QME.
Our route of the formulation and the well-defined approximations allow us to clearly understand the structure of the framework with sufficient generality.
Hence, the approach is immediately applicable to a wide range of physical systems.
As an example, the results are demonstrated by applying the framework to the DLZ model.
Our scheme thus resolves the contentious issue, achieves an extension of the applicable range, and as a result opens up a new avenue for exploring the frontier in driven open quantum systems beyond the adiabatic regime.

The paper is organized as follows.
In Section~\ref{sec:Adiabaticity}, we introduce $\tau_{\A}(t)$ that quantifies the temporal change of $\oH_{\S}(t)$ by assuming the {\em closed system}.
We also discuss the similarities and differences between the adiabaticity of $\oH_{\S}(t)$ and the ordinary adiabatic theorem.
Next, in Section~\ref{sec:QME}, we study the general derivation of the Markovian QME under the WCA.
To help explain the framework clearly, we shall make a brief review of the {\em time-independent} Hamiltonian case.
We then introduce an approximation based on the instantaneous eigenbasis and illustrate the validity beyond the adiabatic regime.
In Section~\ref{sec:DLZ}, we demonstrate the results by applying the framework to the DLZ model.
Finally, in Section~\ref{sec:Conclusions}, we conclude and give an outlook.
Throughout the paper, we set $\hbar=\kB=1$ for simplicity.

\section{Adiabaticity of the system Hamiltonian}\label{sec:Adiabaticity}
We consider an open quantum system with the time-dependent system Hamiltonian $\oH_{\S}(t)$.
Although this topic has been studied by many authors in the past~\cite{Amin08, Childs01, Vega10, Albash12, Davies78, Childs01, Vega10, Cai10, Albash12, Kamleitner13, Pekola10,Vega15,Whitney11, Nalbach14, Javanbakht15}, there is no discussion on how to quantify the temporal change timescale $\tau_{\A}(t)$ of the system Hamiltonian.
Therefore, in this section, we start with the definition of $\tau_{\A}(t)$ by considering the {\em closed system}.
For this purpose, it is convenient to consider a problem about in what condition $\oH_{\S}(t \pm \delta t)$ remains unchanged from $\oH_{\S}(t)$,
\begin{align}
\oH_{\S}(t \pm \delta t) \simeq \oH_{\S}(t),
\label{eq:AH1}
\end{align}
where $\delta t > 0$ is an arbitrary time.
We note that this problem is different from the traditional adiabatic theorem~\cite{Born28,Kato50} in the following sense.
In the adiabatic theorem, the time evolution of the system's state $\ket{\psi_{\S}(t)}$ is considered according to the Schr\"odinger equation $\ii \frac{\dd}{\dd t} \ket{\psi_{\S}(t)}=\oH_{\S}(t) \ket{\psi_{\S}(t)}$.
Then, one finds that the transitions between the instantaneous eigenstates are suppressed when $\oH_{\S}(t)$ changes sufficiently slowly if there is no degeneracy. 
In contrast, in Eq.~\eqref{eq:AH1}, we do {\em not} focus on the time evolution of $\ket{\psi_{\S}(t)}$ but our attention is paid to the time dependence of $\oH_{\S}(t)$ itself.
Therefore, it is always possible to find $\delta t$ that satisfies Eq.~\eqref{eq:AH1}, no matter how fast $\oH_{\S}(t)$ changes when $\oH_{\S}(t)$ is analytic.
In this context, $\delta t$ would be arbitrary as long as it is sufficiently small compared to a certain timescale.
Here, in fact, it is natural to introduce $\tau_\A(t)$ by this certain timescale because, if $\tau_\A(t)$ denotes the temporal change timescale of $\oH_{\S}(t)$, Eq.~\eqref{eq:AH1} should be satisfied when $\delta t$ is much shorter than $\tau_\A(t)$,
\begin{align}
\delta t \ll \tau_{\A}(t).
\label{eq:AH2}
\end{align}
Therefore, we can say that $\oH_{\S}(t)$ is adiabatic, or ``slow'', within the time $\delta t$ when $\delta t$ satisfies the condition of Eq.~\eqref{eq:AH2}.
In this sense, $\tau_{\A}(t)$ would give a clear measure for the adiabaticity of $\oH_{\S}(t)$ itself.

However, Eqs.~\eqref{eq:AH1} and \eqref{eq:AH2} alone do not make much practical sense because we do not know how to estimate $\tau_{\A}(t)$.
As a convenient way to obtain $\tau_\A(t)$, therefore, we use the instantaneous eigenbasis, defined by $\oH_{\S}(t)\ket{n(t)} = \epsilon_n(t) \ket{n(t)}$, where $\ket{n(t)}$ and $\epsilon_n(t)$ denote the $n$-th instantaneous eigenstate and eigenenergy, respectively; $n=0$ labels the ground state.
The states $\ket{n(t)}$ are assumed to be normalized and nondegenerate.
We then read Eq.~\eqref{eq:AH1} in this basis as
\begin{align}
\epsilon_n(t \pm \delta t) \simeq \epsilon_n(t), \quad \ket{n(t \pm \delta t)} \simeq \ket{n(t)},
\label{eq:AES}
\end{align}
for all $n$.
Here, in a similar manner to Eq.~\eqref{eq:AH2}, we can separately introduce $\tau_{\A\E}(t)$ and $\tau_{\A\S}(t)$ that describe the temporal change timescales of the instantaneous eigenenergy and eigenstate.
Hence, the first and second approximations in Eq.~\eqref{eq:AES} are appropriate, respectively, when
\begin{align}
\delta t \ll \tau_{\A\E}(t), \quad \delta t \ll \tau_{\A\S}(t).
\label{eq:tauAES}
\end{align}
As a result, $\tau_\A(t)$ is given by
\begin{align}
\tau_{\A}(t) = \min\{ \tau_{\A\E}(t), \tau_{\A\S}(t) \},
\label{eq:tauA}
\end{align}
because $\epsilon_n(t \pm \delta t) \simeq \epsilon_n(t)$ and $\ket{n(t \pm \delta t)} \simeq \ket{n(t)}$ should both be satisfied for $\delta t \ll \tau_{\A}(t)$.
Since the Taylor expansion of $\epsilon_n(t \pm \delta t)$ up to the first order yields $\epsilon_n(t \pm \delta t) \simeq \epsilon_n(t) \{  1 \pm  \frac{\delta t}{\epsilon_n (t)} \frac {\dd \epsilon_n(t)}{\dd t} \}$, we obtain $\tau_{\A\E}(t)$ as
\begin{align}
\tau_{\A\E}^{-1}(t) &= \max_{n} |\tfrac{1}{\epsilon_n(t)} \tfrac{\dd}{\dd t} \epsilon_n(t)|.
\label{eq:tauAE}
\end{align}
Then, one can indeed obtain $\epsilon_n(t \pm \delta t) \simeq \epsilon_n(t)$ if $\delta t \ll \tau_{\A\E}(t)$.
On the other hand, for $m \ne n$, $|\braket{m(t)|n(t \pm \delta t)}| \ll 1$ should be satisfied when $\ket{n(t \pm \delta t)} \simeq \ket{n(t)}$ because $\braket{m(t)|n(t)}=0$.
Therefore, since $\ket{n(t \pm \delta t)} \simeq \ket{n(t)} \pm \delta t \frac{\dd}{\dd t} \ket{n(t)}$, $\tau_{\A\S}(t)$ is given by
\begin{align}
\tau_{\A\S}^{-1}(t) &= \max_{m \neq n} |\bra{m(t)} \tfrac{\dd}{\dd t} \ket{n(t)}|,
\label{eq:tauAS}
\end{align}
where the maximum is taken over all $m$ and $n$ except for $m=n$.
Thus, $\tau_\A(t)$ can be estimated from Eqs.~\eqref{eq:tauA}--\eqref{eq:tauAS} with the instantaneous eigenbasis.
For the convenience of the reader, the timescales and the situations discussed in this paper are summarized in Tables~\ref{tbl:timescales} and \ref{tbl:conditions}.
While there are a multitude of timescales, we should mention that $\tau_{\A}(t)$, $\tau_{\A\E}(t)$ and  $\tau_{\A\S}(t)$ are the only timescales essentially introduced in this paper.
The other timescales are just straightforward extensions of previously known timescales. 

At this stage, let us discuss the time evolution of $\ket{\psi_{\S}(t)}$.
This is actually equivalent to discuss the evolution operator defined by
\begin{align}
\ou_{\S}(t_2, t_1) \equiv \left \{
	\begin{array}{ll}
	\T \exp \{ -\ii \int^{t_2}_{t_1} \dd s \hat{H}_{\S}(s)\} & \quad t_2 \geq t_1 \\
	\bar{\T} \exp \{ +\ii \int^{t_1}_{t_2} \dd s \hat{H}_{\S}(s)\} & \quad t_1 > t_2 \\
	\end{array}
\right.,
\label{eq:EVS}
\end{align}
where $\T$ ($\bar{\T}$) denotes the chronological (anti-chronological) time ordering operator.
The formal solution of the Schr\"odinger equation is then given by $\ket{\psi_{\S}(t)} = \ou_{\S}(t,t_0)\ket{\psi_{\S}(t_0)}$ with the initial time $t_0$ because $\ou_{\S}(t_2, t_1)$ satisfies
\begin{subequations}
\label{eq:EVprop2}
\begin{align}
&+\ii\textstyle{\frac{\dd}{\dd t_2}} \ou_{\S}(t_2, t_1) = \oH_{\S}(t_2)\ou_{\S}(t_2, t_1), \\
&-\ii\textstyle{\frac{\dd}{\dd t_2}} \ou_{\S}(t_1, t_2) = \ou_{\S}(t_1, t_2)\oH_{\S}(t_2),
\end{align}
\end{subequations}
for $t_2$ later than $t_1$ ($t_2 \ge t_1$).
We note, in addition, that the evolution operator also satisfies
\begin{subequations}
\label{eq:EVprop1}
\begin{align}
&\ou_{\S}(t_3, t_2)\ou_{\S}(t_2, t_1) = \ou_{\S}(t_3, t_1), \\
&\ou_{\S}(t_2, t_1) = \du_{\S}(t_1, t_2) = \ou^{-1}_{\S}(t_1, t_2),
\end{align}
\end{subequations}
by definition.
Then, if $\delta t > 0$ satisfies Eq.~\eqref{eq:AH2}, the evolution operator from $t - \delta t$ to $t$, i.e.~$\ou_{\S}(t,t-\delta t)$, can be approximated by
\begin{align}
\ou_{\S}(t,t-\delta t) &= \exp \left\{ -\ii \textstyle{\int^{\delta t}_{0}} \dd s' \oH_{\S}(t-s') \right\} \nonumber \\
& \simeq e^{-\ii \oH_{\S}(t) \delta t},
\label{eq:EV(t;t-delta t)}
\end{align}
where the integral variable is changed to $s' = t - s$ [see also Eq.~\eqref{eq:EVS}] in the first line and Eq.~\eqref{eq:AH1} has been used in the second line.
This equation means that the evolution of the eigenstates can be approximated just by the dynamical phase shift, 
\begin{align}
\ou_{\S}(t,t-\delta t) &\simeq \sum_{n} e^{-\ii \epsilon_n(t) \delta t} \ket{n(t)}\bra{n(t)},
\label{eq:EV(t;t-delta t)2}
\end{align}
when $\delta t$ is much shorter than $\tau_{\A}(t)$.
These discussions based on the concept of the adiabaticity of $\oH_{\S}(t)$ will play an important role when we develop the Markovian QME with $\oH_{\S}(t)$ in the following sections.

Here, it would be instructive to see the similarities and differences between the adiabaticity of $\oH_{\S}(t)$ and the ordinary adiabatic theorem.
For this purpose, we introduce an {\em intrinsic evolution timescale} of $\oH_{\S}(t)$ by
\begin{align}
\tau_{\E}^{-1}(t) \equiv \min_{m \ne n} |\epsilon_{mn}(t)|,
\label{eq:tauE}
\end{align}
where $\epsilon_{mn}(t) \equiv \epsilon_m(t) - \epsilon_n(t)$ denotes the instantaneous energy gap.
Hence, $\tau_{\E}(t)$ also corresponds to the longest timescale of the intrinsic oscillation period in the off-diagonal density matrix elements.  
One can then easily confirm that the adiabatic theorem is validated if 
\begin{align}
\tau_{\E}(t) \ll \tau_{\A\S}(t),
\label{eq:AT}
\end{align}
is satisfied for all relevant time $t$ (see Ref.~\onlinecite{Sarandy2004}, for example) by using $\bra{m(t)} \tfrac{\dd}{\dd t} \ket{n(t)} = \bra{m(t)} \tfrac{\dd \oH_\S(t)}{\dd t} \ket{n(t)}/\epsilon_{nm}(t)$.
Equation~\eqref{eq:AT} means that all the eigenstates $\ket{n(t)}$ should be adiabatic with respect to $\tau_\E(t)$.
In other words, the eigenstates $\ket{n(t)}$ should remain unchanged at least within the intrinsic oscillation period $\tau_\E(t)$ to satisfy the adiabatic theorem.
In that case, the evolution operator $\ou_{\S}(t_2, t_1)$ can be approximated by $\ou_{\S}(t_2, t_1) \simeq \ou^{\ad}_{\S}(t_2, t_1)$ with~\cite{Mostafazadeh97,Albash12}
\begin{align}
\ou^{\ad}_{\S}(t_2, t_1) \equiv \sum_{n} e^{-\ii \mu_n(t_2, t_1)} \ket{n(t_2)}\bra{n(t_1)},
\label{eq:EVad}
\end{align}
where $\mu_n(t_2, t_1) \equiv \int^{t_2}_{t_1} \dd s \{ \epsilon_n(s) - \phi_n(s) \}$ and $\phi_n(s) \equiv \ii \bra{n(s)}\frac{\dd}{\dd s}\ket{n(s)}$ denotes the Berry connection. 
Notice that, according to Eq.~\eqref{eq:EVad}, the $n$-th eigenstate $\ket{n(t_1)}$ at time $t_1$ will evolve to the $n$-th eigenstate $\ket{n(t_2)}$ at time $t_2$ with the phase shift $\mu_n(t_2, t_1)$.
Therefore, Eq.~\eqref{eq:EVad} directly means that there are no transitions between the instantaneous eigenstates.
We refer to this type of dynamics as the {\em adiabatic evolution}.
By comparing Eq.~\eqref{eq:AH2} with Eq.~\eqref{eq:AT} [Eq.~\eqref{eq:EV(t;t-delta t)2} with Eq.~\eqref{eq:EVad}], the adiabaticity of $\oH_{\S}(t)$ is formally similar to the adiabatic theorem.
However, they are conceptually different from each other, as described below Eq.~\eqref{eq:AH1}.
According to Eq.~\eqref{eq:AH2}, we can indeed always find $\delta t$ even if $\tau_{\A}(t)$ becomes short due to a rapid change of $\oH_{\S}(t)$.
As a result, Eq.~\eqref{eq:EV(t;t-delta t)2} is validated, in which the eigenstates are left unchanged except for their phase factors.
In contrast, Eq.~\eqref{eq:AT} fails when $\tau_{\A\S}(t)$ becomes short due to the rapid change of $\oH_{\S}(t)$.
This is because $\tau_{\E}(t)$ is fixed once the structure of $\oH_{\S}(t)$ is determined.
Therefore, the adiabatic evolution [Eq.~\eqref{eq:EVad}] is valid only when $\oH_{\S}(t)$ varies slowly to satisfy Eq.~\eqref{eq:AT}.
With the help of this distinction, we shall see below that the Markovian QME is naturally derived without the adiabatic theorem.

\begin{table*}[!tb]
\caption{\label{tbl:timescales}Definitions of typical timescales.
Here, $\epsilon_n(t)$ and $\ket{n(t)}$ denote the instantaneous eigenenergy and eigenstate of $\oH_{\S}(t)$, respectively.
$\epsilon(t)$ and $\epsilon'(t)$ are the Bohr frequencies of the system.
$\gamma_{\alpha\beta}(\omega)$ is the Fourier transform of the bath correlation function $C_{\alpha\beta}(\tau)$.
$\gamma_{\peak,\alpha\beta}$ is the peak value of $\gamma_{\alpha\beta}(\omega)$.
See also the text for details.}
\begin{ruledtabular}
\begin{tabular}{clll}
$\tau_{\A}(t)$ & Temporal change timescale of $\oH_{\S}(t)$ & Eq.~\eqref{eq:tauA} & $\tau_{\A}(t) \equiv \min\{ \tau_{\A\E}(t), \tau_{\A\S}(t) \}$ \\
$\tau_{\A\E}(t)$ & Temporal change timescale of $\epsilon_n(t)$ & Eq.~\eqref{eq:tauAE} & $\tau_{\A\E}^{-1}(t) \equiv \max_{n} |\tfrac{1}{\epsilon_n(t)} \tfrac{\dd}{\dd t} \epsilon_n(t)|$ \\
$\tau_{\A\S}(t)$ & Temporal change timescale of $\ket{n(t)}$ & Eq.~\eqref{eq:tauAS} & $\tau_{\A\S}^{-1}(t) \equiv \max_{m \neq n} |\bra{m(t)} \tfrac{\dd}{\dd t} \ket{n(t)}|$ \\
$\tau_{\E}(t)$ & Intrinsic evolution timescale & Eq.~\eqref{eq:tauE} & $\tau_{\E}^{-1}(t) \equiv \min_{m \ne n} |\epsilon_m(t) - \epsilon_n(t)|$\\
$\tau_{\R}(t)$ & Relaxation timescale & Eq.~\eqref{eq:IEA::tauR} & $\tau_{\R}^{-1}(t) \equiv \max_{\alpha, \beta, \epsilon(t)} \gamma_{\alpha\beta}(\epsilon(t))$\\
$\tau_{\S}(t)$ & Intrinsic beat timescale & Eq.~\eqref{eq:IEA::tauS} & $\tau_{\S}^{-1}(t) \equiv \min_{\epsilon(t) \ne \epsilon'(t)} |\epsilon(t)-\epsilon'(t)|$ \\
$\tau_{\R}^{\text{min}}$ & Minimum relaxation timescale & Eq.~\eqref{eq:tauRmin} & $(\tau_{\R}^{\text{min}})^{-1} \equiv \max_{\alpha,\beta} \gamma_{\peak,\alpha\beta}$\\
$\tau_{\B}$ & Decay timescale of the bath correlation & & \\
$\varDelta t \equiv t-t_0$ & Elapsed time from the initial time $t_0$ & & \\
\end{tabular}
\end{ruledtabular}

\caption{\label{tbl:conditions} Summary of the situations; see also the text for details.}
\begin{ruledtabular}
\begin{tabular}{clll}
Adiabatic evolution & Eq.~\eqref{eq:AT} & $ \tau_{\E}(t) \ll \tau_{\A\S}(t) $ &\\
Weak-coupling approximation (WCA) & Eq.~\eqref{eq:WCAcondition} & $ \tau_{\B} \ll \tau_{\R}^{\text{min}} $ & \multirow{2}{*}{ $\biggr\}$ Born-Markov approximation}\\
Markov approximation & Eq.~\eqref{eq:MA} & $ \tau_{\B} \ll \varDelta t $ \\
Instantaneous eigenbasis approximation (IEA) & Eq.~\eqref{eq:IEAcondition} & $ \tau_{\B} \ll \tau_{\A}(t) $ & \\
Secular approximation (SA) & Eq.~\eqref{eq:IEA::SA} & \multicolumn{2}{l}{ $\tau_{\S}(t) \ll \tau_{\R}(t)$ and $\tau_{\S}(t) \ll \tau_{\A}(t)$} \\
Neglect of relaxation & Eq.~\eqref{eq:NeglectRelaxation} & $ \tau_{\A}(t) \ll \tau_{\text{R}}(t) $ & \\
\end{tabular}
\end{ruledtabular}
\end{table*}

\section{Quantum Master Equation}\label{sec:QME}
\subsection{Weak-coupling and Markovian Approximation}\label{subsec:WCMA}
We now turn to the study of the Markovian QME under the WCA.
Here, in order to  fix our notations as well as to give a self-contained presentation, let us shortly review the standard steps to obtain the Markovian QME.
In the Schr\"odinger picture, the general Hamiltonian we consider is 
\begin{align*}
\hat{H}(t) &= \oH_{0}(t) + \oH_{\S\B} = \oH_{\S}(t) + \oH_{\B} + \oH_{\S\B},
\end{align*}
where $\oH_{\B}$ is the Hamiltonian of the bath and 
\begin{align}
\oH_{\S\B} \equiv \sum_{\alpha} \oS_{\alpha} \otimes \oB_{\alpha},
\label{eq:HSB}
\end{align}
is the system-bath interaction Hamiltonian.
Without loss of generality, we assume that $\oS_\alpha$ and $\oB_\alpha$ are the Hermitian operators acting only on the system and bath Hilbert spaces, respectively~\cite{Rivas12}.
The total density operator $\oRho(t)$ evolves according to the von Neumann equation; $\frac{\dd}{\dd t}\oRho(t) = -\ii [\oH(t), \oRho(t)]$.

Let $\ou_{0}(t_2, t_1)$ describe the evolution operator for $\oH_{0}(t)$ that is defined in the same way as Eq.~\eqref{eq:EVS} except for the replacement $\oH_{\S}(t) \to \oH_{0}(t)$.
We then transform into the interaction picture with respect to $\oH_{0}(t)$ by introducing
\begin{align*}
\vO(t) \equiv \du_0(t, t_0)\oO(t)\ou_0(t, t_0),
\end{align*}
where $\oO(t)$ is an arbitrary operator.
In the interaction picture, therefore, the von Neumann equation is written as $\frac{\dd}{\dd t}\vRho(t) = -\ii [\vH_{\S\B}(t), \vRho(t)]$.
As a result, the standard time-convolutionless technique using projection superoperators~\cite{Breuer02,Rivas12} yields the equation of motion for the reduced density operator $\vRho_{\S}(t) \equiv \TrB{\vRho(t)}$,
\begin{align}
\tfrac{\dd}{\dd t} \vRho_{\S}(t) &= - \textstyle{ \int_{t_0}^{t} \dd s} \TrB{ \vH_{\S\B}(t), [\vH_{\S\B}(s), \vRho_{\S}(t) \otimes \oRho_{\B}] },
\label{eq:nMRedfield1}
\end{align}
up to the second order in $\oH_{\S\B}$, or equivalently within the WCA; see also Appendix~\ref{TCL}.
In the derivation, we have assumed that the initial state is separable, $\oRho(t_0) = \oRho_{\S}(t_0) \otimes \oRho_{\B}$, and that the odd moments of $\vH_{\S\B}(t)$ with respect to $\oRho_{\B}$ vanish;
\begin{align}
\TrB{\vH_{\S\B}(t_1)\vH_{\S\B}(t_2) \cdots \vH_{\S\B}(t_{2n+1}) \oRho_{\B}}=0,
\label{eq:OddMoments}
\end{align}
where $n = 0, 1, 2, \cdots$.
By inserting $\vH_{\S\B}(t) = \sum_{\alpha} \vS_{\alpha}(t) \otimes \vB_{\alpha}(t)$ into Eq.~\eqref{eq:nMRedfield1}, with a change of the integration variable to $\tau = t - s$, we obtain
\begin{align}
\tfrac{\dd}{\dd t} \vRho_{\S}(t) =& \textstyle{\sum_{\alpha, \beta} \int^{\varDelta t}_{0}\dd \tau} C_{\alpha\beta}(\tau)
 \{ \vS_{\beta}(t-\tau) \vRho_{\S}(t) \dvS_{\alpha}(t) \nonumber\\
&- \dvS_{\alpha}(t)\vS_{\beta}(t-\tau) \vRho_{\S}(t)\} + \Hc,
\label{eq:nMRedfield2}
\end{align}
where $\varDelta t \equiv t - t_0$ denotes the elapsed time from the initial time $t_0$ and $C_{\alpha\beta}(\tau) \equiv \TrB{\vB_{\alpha}(\tau) \vB_{\beta}(0) \oRho_{\B}}$ is the bath correlation function.
The typical decay timescale of $C_{\alpha\beta}(\tau) $ will be denoted by $\tau_{\B,\alpha\beta}$.
In the derivation of Eq.~\eqref{eq:nMRedfield2}, we have assumed that $\oRho_{\B}$ is in the steady state, i.e. $[\oH_{\B}, \oRho_{\B}] = 0$, to obtain $\TrB{\vB_{\alpha}(t)\vB_{\beta}(t-\tau) \oRho_{\B} } = \TrB{\vB_{\alpha}(\tau) \vB_{\beta}(0) \oRho_{\B}}$ for simplicity.
The thermal bath with the temperature $T$ is the most typical example; $\oRho_{\B} = \exp (-\oH_{\B}/T)/\TrB{\exp (-\oH_{\B}/T)}$.
We remark that Eqs.~\eqref{eq:nMRedfield1} and \eqref{eq:nMRedfield2} are time-convolutionless in the sense that there is no time convolution in terms of $\vRho_{\S}(t)$.
However, the time convolution between $C_{\alpha\beta}(\tau)$ and $\vS_{\beta}(t-\tau)$ is still required, which will cause a practical difficulty when we consider the time-dependent Hamiltonian explicitly.

{\em The Markovian approximation}---Here, in Eq.~\eqref{eq:nMRedfield2}, $C_{\alpha\beta}(\tau)$ decays to zero well after the decay timescale of the bath correlation function $\tau_{\B} \equiv \max_{\alpha, \beta} \tau_{\B,\alpha\beta}$.
Therefore, the upper limit of the integral may be approximated by infinity if we are only interested in the system dynamics over the time which is longer than  $\tau_{\B}$;
\begin{align}
\tau_{\B} \ll \varDelta t.
\label{eq:MA}
\end{align}
Under this approximation, known as the Markovian approximation~\cite{Note1}, we finally obtain
\begin{align}
\tfrac{\dd}{\dd t} \vRho_{\S}(t) =& \textstyle{\sum_{\alpha, \beta} \int^{\infty}_{0}\dd \tau} C_{\alpha\beta}(\tau)
 \{ \vS_{\beta}(t-\tau) \vRho_{\S}(t) \dvS_{\alpha}(t) \nonumber\\
&- \dvS_{\alpha}(t)\vS_{\beta}(t-\tau) \vRho_{\S}(t)\} + \Hc.
\label{eq:MQME}
\end{align}
We note however that, in the vicinity of the initial time, $\varDelta t \lesssim \tau_{\B}$, the Markovian approximation is not validated because $C_{\alpha\beta}(\tau)$ does not yet sufficiently decay.
As a result, errors accumulated in $\varDelta t \lesssim \tau_{\B}$ sometimes largely affect the subsequent evolution even for $\varDelta t \gg \tau_{\B}$.
This problem is nevertheless avoidable by using the renormalized or ``slipped" initial condition~\cite{Suarez92,Gaspard99} (see also Appendix~\ref{SLP}).
Therefore, in the followings, we limit ourselves only to the Markovian QME under the WCA, which is equivalent to the QME under the Born-Markov approximation~\cite{Breuer02}. 
We do not go into the non-Markovian dynamics recently highlighted by several authors~\cite{Wolf08, Breuer09, Rivas10, Diosi14, Ferialdi16, Breuer16} but the concept of the Markovianity is used in the same spirit that there is no memory effect coming back from the bath~\cite{Breuer16}.

{\em Time-independent case}---Equation~\eqref{eq:MQME} can be rewritten in a more tractable form when the system Hamiltonian is time-independent; $\oH_{\S}(t) = \oH_{\S}$, $\ket{n(t)} = \ket{n}$, and $\epsilon_n(t) = \epsilon_n$.
To see this, let us define
\begin{align}
\oS_{\beta}(\epsilon) &\equiv \sum_{\epsilon_{mn}=\epsilon} \oPi(\epsilon_n) \oS_{\beta} \oPi(\epsilon_m), 
\label{eq:Dcmp} \\
\oPi(\epsilon_n) &\equiv \ket{n}\bra{n},
\label{eq:Prjc}
\end{align}
where the summation in Eq.~\eqref{eq:Dcmp} is taken over all $m$ and $n$ satisfying $\epsilon_m -\epsilon_n = \epsilon$.
Then, the summation over all possible $\epsilon$ (the Bohr frequencies of the system) yields the spectral decomposition, 
\begin{align}
\oS_{\beta} = \sum_{\epsilon} \oS_{\beta}(\epsilon),
\label{eq:Spctrl-Dcmp}
\end{align}
due to the completeness of the eigenbasis.
Since the evolution operator $\ou_{\S}(t, t_0)$ [Eq.~\eqref{eq:EVS}] is simply described by $\ou_{\S}(t, t_0) = e^{ -\ii (t-t_0)\oH_{\S} }$ for $\oH_{\S}(t) = \oH_{\S}$, $\vS_{\beta}(t-\tau)$ in Eq.~\eqref{eq:MQME} is given by
\begin{align}
\vS_{\beta}(t-\tau) = \sum_{\epsilon} e^{ \ii \epsilon \tau} \vS_{\beta}(\epsilon; t),
\label{eq:tau-separation}
\end{align}
where $\vS_{\beta}(\epsilon; t)$ is the interaction picture of $\oS_{\beta}(\epsilon)$; $\vS_{\beta}(\epsilon; t) \equiv e^{\ii (t-t_0)\oH_{\S}} \oS_{\beta}(\epsilon) e^{-\ii (t-t_0)\oH_{\S}} = e^{-\ii (t-t_0)\epsilon}  \oS_{\beta}(\epsilon)$.
Note that the key point in Eq.~\eqref{eq:tau-separation} is that the $\tau$-dependence of $\oS_{\beta}(t-\tau)$ is separated from the operator part.
Therefore, by inserting Eq.~\eqref{eq:tau-separation} into Eq.~\eqref{eq:MQME}, the time convolution between $C_{\alpha\beta}(\tau)$ and $\vS_{\beta}(t-\tau)$ yields
\begin{align}
\int^{\infty}_{0} \dd \tau C_{\alpha\beta}(\tau) \vS_{\beta}(t-\tau) 
= \sum_{\epsilon} \varGamma_{\alpha\beta}(\epsilon) \vS_{\beta}(\epsilon; t),
\label{eq:CT}
\end{align}
where $\varGamma_{\alpha\beta}(\epsilon)$ denotes the one-sided Fourier transform of the bath correlation function
\begin{align}
\varGamma_{\alpha\beta}(\epsilon) \equiv \textstyle{\int^{\infty}_{0}\dd \tau} e^{ \ii \epsilon \tau} C_{\alpha\beta}(\tau).
\label{eq:Gamma}
\end{align}
Notice that this is an analogy to the ordinary convolution theorem, i.e.~the convolution in time domain is equivalent to the multiplication in the Fourier domain.
As a result, one obtains the standard form of the QME in the interaction picture,
\begin{align}
\tfrac{\dd}{\dd t} \vRho_{\S}(t) =& \vD_t \vRho_{\S}(t),
\label{eq:std::QME}
\end{align}
where the dissipator $\vD_t$ is defined by
\begin{multline}
\vD_t \vRho_{\S} \equiv \textstyle{\sum_{\epsilon} \sum_{\alpha, \beta} } \varGamma_{\alpha\beta}(\epsilon) 
 \{ \vS_{\beta}(\epsilon; t) \vRho_{\S} \dvS_{\alpha}(t) \\
- \dvS_{\alpha}(t)  \vS_{\beta}(\epsilon; t) \vRho_{\S}\} + \Hc.
\label{eq:std::D}
\end{multline}
However, the dissipator $\vD_t$ is still not described by the Lindblad form and this means that the complete positivity is not guaranteed.
Therefore, the secular approximation (SA), or equivalently the rotating wave approximation (RWA), is often applied, where rapidly oscillating terms in Eq.~\eqref{eq:std::D} is neglected (averaged out to zero).
This is performed by putting $\dvS_{\alpha}(t) = \sum_{\epsilon'}\dvS_{\alpha}(\epsilon'; t)$ [cf. Eq.~\eqref{eq:tau-separation}] into Eq.~\eqref{eq:std::D} and by neglecting all terms with $\epsilon \ne \epsilon'$.
We note that this approximation is based on the oscillating forms of $\dvS_{\alpha}(\epsilon; t) = e^{\ii (t-t_0)\epsilon}  \dS_{\alpha}(\epsilon)$ and $\vS_{\beta}(\epsilon; t) = e^{-\ii (t-t_0)\epsilon}  \oS_{\beta}(\epsilon)$.
Then, we finally arrive at $\vD_t \simeq \vD^{\S\A}_t$ with
\begin{multline}
\vD^{\S\A}_t \vRho_{\S} \equiv \textstyle{\sum_{\epsilon} \sum_{\alpha, \beta} } \varGamma_{\alpha\beta}(\epsilon) 
 \{ \vS_{\beta}(\epsilon; t) \vRho_{\S} \dvS_{\alpha}(\epsilon; t) \\
- \dvS_{\alpha}(\epsilon; t)  \vS_{\beta}(\epsilon; t) \vRho_{\S}\} + \Hc,
\label{eq:std::DSA}
\end{multline}
which indeed results in the Lindblad form.
Note however that the SA is justified only when all the differences of the Bohr frequencies, i.e. $|\epsilon - \epsilon'|$ for $\epsilon \ne \epsilon'$, are larger than the relaxation rate of the system~\cite{Breuer02}.
On the one hand, the relaxation rate of the system $\tau^{-1}_{\R}$ is characterized by
\begin{align}
\tau_{\R}^{-1} = \max_{\alpha, \beta, \epsilon} \gamma_{\alpha\beta}(\epsilon),
\label{eq:std::tauR}
\end{align}
where $\gamma_{\alpha\beta}(\epsilon) \equiv \varGamma_{\alpha\beta}(\epsilon) +  \varGamma_{\beta\alpha}^{*}(\epsilon) = \int^{\infty}_{-\infty} \dd \tau e^{\ii\epsilon\tau} C_{\alpha\beta}(\tau)$.
On the other hand, we can define an intrinsic timscale $\tau_{\S}$ by using the minimum of $|\epsilon - \epsilon'|$ for $\epsilon \ne \epsilon'$; 
\begin{align}
\tau_{\S}^{-1} \equiv \min_{\epsilon \ne \epsilon'} |\epsilon-\epsilon'|.
\label{eq:std::tauS}
\end{align}
Therefore, the condition for the SA is described by $\tau^{-1}_{\R}  \ll \tau^{-1}_{\S}$, or equivalently,
\begin{align}
\tau_{\S}  \ll \tau_{\R}.
\label{eq:std::SAcondition}
\end{align}
Here, we refer to $\tau_{\S}$ as the intrinsic beat timescale to distinguish from Eq.~\eqref{eq:tauE}, while $\tau_{\S}$ is also called the intrinsic evolution timescale~\cite{Breuer02}.
This is because the differences of the Bohr frequencies correspond to the beat frequencies of the intrinsic oscillation of the off-diagonal density matrix elements.
If the condition is not satisfied, the SA sometimes leads to unphysical results and one often has to use $\vD_t$ or alternatively approximated Lindblad QME~\cite{Wichterich07, Gaspard99, Esposito09, Yuge15}. 

\subsection{Instantaneous Eigenbasis Approximation}\label{subsec:IEA}
Now, we return to the problem of the time-dependent system Hamiltonian, $\oH_{\S}(t)$.
In this case, our starting point is Eq.~\eqref{eq:MQME}.
Although this equation can be solved numerically in principle, the numerical cost would not be low in practice because Eq.~\eqref{eq:MQME} requires the numerical $\tau$-integration at each time step.
It is therefore desirable to analytically reduce Eq.~\eqref{eq:MQME} into a similar form to $\vD_t$ [Eq.~\eqref{eq:std::D}] or $\vD^{\S\A}_t$ [Eq.~\eqref{eq:std::DSA}].
In this context, one often assumes that $\oH_{\S}(t)$ is slowly varying, and then, replaces the Bohr frequencies $\epsilon$, the eigenenergies $\epsilon_n$, and the eigenstates $\ket{n}$ in the dissipators by the corresponding time-dependent ones,  $\epsilon(t)$, $\epsilon_n(t)$, and $\ket{n(t)}$, respectively.
Indeed, such an approach has been employed by Childs {\it et al}.~\cite{Childs01} and its microscopic derivation has recently been shown by Albash {\it et al}., based on the `ideal' adiabatic evolution operator $\ou^{\ad}_{\S}(t_2, t_1)$ [Eq.~\eqref{eq:EVad}], or equivalently, the adiabatic theorem~\cite{Albash12}.
In contrast, our purpose below is to derive such dissipators {\em without using the adiabatic theorem}.

To this end, we first introduce 
\begin{align}
\oS_{\beta}(\epsilon(t)) &= \sum_{\epsilon_{mn}(t)=\epsilon(t)} \oPi(\epsilon_n(t)) \oS_{\beta} \oPi(\epsilon_m(t)), 
\label{eq:Dcmp2} \\
\oPi(\epsilon_n(t)) &= \ket{n(t)}\bra{n(t)},
\label{eq:Prjc2}
\end{align}
in the same manner as Eqs.~\eqref{eq:Dcmp} and \eqref{eq:Prjc}.
Therefore, the summation over all possible $\epsilon(t)$ yields the decomposition similar to Eq.~\eqref{eq:Spctrl-Dcmp},
\begin{align}
\oS_{\beta} = \sum_{\epsilon(t)} \oS_{\beta}(\epsilon(t)).
\label{eq:Spctrl-Dcmp2}
\end{align}
again due to the completeness of the (instantaneous) eigenbasis.
As illustrated in the previous section, the advantage to use such a decomposition was to separate the $\tau$-dependence from $\vS_{\beta}(t-\tau)$, based on the explicit calculation of the time dependence in the interaction picture [see Eq.~\eqref{eq:tau-separation}].
Then, the time convolution between $C_{\alpha\beta}(\tau)$ and $\vS_{\beta}(t-\tau)$ becomes possible [Eq.~\eqref{eq:CT}].
However, this approach is now {\em not} allowed for the time-dependent Hamiltonian $\oH_{\S}(t)$ because the evolution operator $\ou_{\S}(t, t_0)$ includes the time integration of $\oH_{\S}(t)$ [Eq.~\eqref{eq:EVS}].

In order to avoid this difficulty, according to Eq.~\eqref{eq:EVprop1}, we rewrite $\vS_{\beta}(t-\tau)$ in the following form, 
\begin{align}
\vS_{\beta}(t-\tau)=\du_{\S}(t,t_0)\ou_{\S}(t,t-\tau)\oS_{\beta}\du_{\S}(t,t-\tau)\ou_{\S}(t,t_0),
\label{eq:EVseparation}
\end{align}
where the evolution operator $\ou_{\S}(t-\tau, t_0)$ has been separated into two parts; $\ou_{\S}(t-\tau, t_0) = \du_{\S}(t,t-\tau)\ou_{\S}(t,t_0)$.
Since $\tau$ is positive in Eq.~\eqref{eq:MQME}, the form of $\ou_{\S}(t,t-\tau)$ is reminiscent of Eq.~\eqref{eq:EV(t;t-delta t)}.
By considering that the $\tau$-integration in Eq.~\eqref{eq:MQME} converges for $\tau \simeq \tau_{\B}$ due to the decay of $C_{\alpha\beta}(\tau)$, we can indeed approximate $\ou_{\S}(t,t-\tau)$ by
\begin{align}
\ou_{\S}(t,t-\tau) \simeq e^{-\ii\oH_{\S}(t)\tau},
\label{eq:EVapprox}
\end{align}
when
\begin{align}
\tau_{\B} \ll \tau_{\A}(t).
\label{eq:IEAcondition}
\end{align}
As a result, Eq.~\eqref{eq:EVseparation} yields 
\begin{align}
\vS_{\beta}(t-\tau) &\simeq \du_{\S}(t, t_0) \left\{ \sum_{\epsilon(t)} e^{\ii\epsilon(t)\tau} \oS_{\beta}(\epsilon(t)) \right\}\ou_{\S}(t, t_0), \nonumber \\
& = \sum_{\epsilon(t)}  e^{\ii\epsilon(t)\tau}  \vS_{\beta}(\epsilon(t); t),
\label{eq:IEA}
\end{align}
where $\vS_{\beta}(\epsilon(t); t)$ is the interaction picture of $\oS_{\beta}(\epsilon(t))$, $\vS_{\beta}(\epsilon(t); t)=\du_{\S}(t,t_0)\oS_{\beta}(\epsilon(t))\ou_{\S}(t,t_0)$, and we have used Eqs.~\eqref{eq:Dcmp2}--\eqref{eq:Spctrl-Dcmp2}.
It is important here to notice that the $\tau$-dependence is separated from the operator part and Eq.~\eqref{eq:IEA} corresponds to a straightforward extension of Eq.~\eqref{eq:tau-separation}.
We refer to Eq.~\eqref{eq:IEA} as the instantaneous eigenbasis approximation (IEA).
Note that the IEA becomes exact and recovers Eq.~\eqref{eq:tau-separation} when $\oH_{\S}(t)$ is time-independent.
This is consistent with the condition for the IEA [Eq.~\eqref{eq:IEAcondition}] because $\tau_{\A}(t)$ goes to infinity based on  Eqs.~\eqref{eq:tauA}--\eqref{eq:tauAS} when $\oH_{\S}(t) = \oH_{\S}$.

By applying the IEA [Eq.~\eqref{eq:IEA}], we can then perform the same steps as Eqs.~\eqref{eq:CT}--\eqref{eq:std::D} to obtain the QME in the interaction picture.
The result is 
\begin{align}
\tfrac{\dd}{\dd t} \vRho_{\S}(t) =& \vD^{\I\E\A}_t \vRho_{\S}(t),
\label{eq:IP::QMEIEA}
\end{align}
where the dissipator $\vD^{\I\E\A}_t$ is defined by
\begin{multline}
\vD^{\I\E\A}_t\oRho_{\S} \equiv \sum_{\epsilon(t)} \sum_{\alpha, \beta} \varGamma_{\alpha\beta}(\epsilon(t)) 
 \{ \vS_{\beta}(\epsilon(t); t) \vRho_{\S} \dvS_{\alpha}(t) \\
- \dvS_{\alpha}(t)  \vS_{\beta}(\epsilon(t); t) \vRho_{\S}\} + \Hc.
\label{eq:IP::DIEA}
\end{multline}
These equations are formally the same as Eq.~\eqref{eq:std::D} except that $\epsilon \to \epsilon(t)$, $\epsilon_n \to \epsilon_n(t)$, and $\ket{n} \to \ket{n(t)}$.
It follows that Eqs.~\eqref{eq:IP::QMEIEA} and \eqref{eq:IP::DIEA} do not guarantee the complete positivity due to the non-Lindblad form.
Therefore, the SA is again required to obtain the Lindblad form in a similar manner to the time-independent case.
However, we remark that $\vS_{\beta}(\epsilon(t); t) \ne e^{-\ii (t-t_0)\epsilon(t)}  \oS_{\beta}(\epsilon(t))$ in $\vD^{\I\E\A}_t$ even though $\vS_{\beta}(\epsilon; t) = e^{-\ii (t-t_0)\epsilon}  \oS_{\beta}(\epsilon)$ holds true in $\vD_t$ because the interaction picture cannot be explicitly obtained when the system Hamiltonian is time-dependent.
Actually, this difference between $\vD^{\I\E\A}_t$ and $\vD_t$ makes the application of the SA more difficult than the time-independent case since the SA is based on the oscillating forms of $\dvS_{\alpha}(\epsilon; t)$ and $\vS_{\beta}(\epsilon; t)$, as seen above Eq.~\eqref{eq:std::DSA}.
Nevertheless, we can show that the SA is still possible (see also Appendix~\ref{sec:SA}) when 
\begin{align}
\tau_{\S}(t) \ll \tau_{\R}(t) \quad \text{and} \quad \tau_{\S}(t) \ll \tau_{\A}(t),
\label{eq:IEA::SA}
\end{align}
are simultaneously satisfied.
Here, $\tau_{\R}(t)$ and $\tau_{\S}(t)$ are the straightforward extensions of Eqs.~\eqref{eq:std::tauR} and \eqref{eq:std::tauS},
\begin{align}
\tau_{\R}^{-1}(t) &\equiv \max_{\alpha, \beta, \epsilon(t)} \gamma_{\alpha\beta}(\epsilon(t)),
\label{eq:IEA::tauR}\\
\tau_{\S}^{-1}(t) &\equiv \min_{\epsilon(t) \ne \epsilon'(t)} |\epsilon(t)-\epsilon'(t)|,
\label{eq:IEA::tauS}
\end{align}
where $\tau_{\R}(t)$ and $\tau_{\S}(t)$ denote the characteristic relaxation timescale of $\vRho_{\S}(t)$ and the intrinsic beat timescale between the different Bohr frequencies $\epsilon(t)$ and $\epsilon'(t)$, respectively.
Physically speaking, therefore, Eq.~\eqref{eq:IEA::SA} means that $\vRho_{\S}(t)$ and $\oH_{\S}(t)$ must remain unchanged until the beat oscillations of $\dvS_{\alpha}(\epsilon; t)$ and $\vS_{\beta}(\epsilon; t)$ are sufficiently developed.
Hence, Eq.~\eqref{eq:IEA::SA} is a natural extension of Eq.~\eqref{eq:std::SAcondition} in the spirit of the SA.
Under this condition, we can obtain $\vD^{\I\E\A}_t \simeq \vD^{\I\E\S\A}_t$ with
\begin{multline}
\vD^{\I\E\S\A}_t\oRho_{\S} \equiv \sum_{\epsilon(t)} \sum_{\alpha, \beta} \varGamma_{\alpha\beta}(\epsilon(t)) 
 \{ \vS_{\beta}(\epsilon(t); t) \vRho_{\S} \dvS_{\alpha}(\epsilon(t); t) \\
- \dvS_{\alpha}(\epsilon(t); t)  \vS_{\beta}(\epsilon(t); t) \vRho_{\S}\} + \Hc,
\label{eq:IP::DIESA}
\end{multline}
which is again formally the same as $\vD^{\S\A}_t$ [Eq.~\eqref{eq:std::DSA}].
Thus, the dissipator can be described in the Lindblad form.

Finally, for practical use, it is better to transform back into the Schr\"odinger picture.
With the help of Eq.~\eqref{eq:EVprop2}, Eq.~\eqref{eq:IP::QMEIEA} gives 
\begin{align}
\tfrac{\dd}{\dd t}\oRho_{\S}(t) &= -\ii [\oH_{\S}(t), \oRho_{\S}(t)] + \oD^{\I\E\A}_t\oRho_{\S}(t),
\label{eq:SP::QMEIEA}
\end{align}
where the dissipator $\oD^{\I\E\A}_t$ is
\begin{align}
\oD^{\I\E\A}_t\oRho_{\S} &= \textstyle{\sum_{\epsilon(t)}\sum_{\alpha,\beta}} \varGamma_{\alpha\beta}(\epsilon(t)) \nonumber\\ 
&\times \{ \oS_{\beta}(\epsilon(t))\oRho_{\S}\dS_{\alpha} - \dS_{\alpha} \oS_{\beta}(\epsilon(t))\oRho_{\S}  \} + \Hc.
\label{eq:SP::DIEA}
\end{align}
After the application of the SA, $\oD^{\I\E\A}_t \simeq \oD^{\I\E\S\A}_t$ is justified with
\begin{multline}
\oD^{\I\E\S\A}_t\oRho_{\S} = \textstyle{\sum_{\epsilon(t)}\sum_{\alpha,\beta}} \varGamma_{\alpha\beta}(\epsilon(t)) \\ 
\times \{ \oS_{\beta}(\epsilon(t))\oRho_{\S}\dS_{\alpha}(\epsilon(t)) - \dS_{\alpha}(\epsilon(t)) \oS_{\beta}(\epsilon(t))\oRho_{\S}  \} + \Hc,
\label{eq:SP::DIESA}
\end{multline}
in the Schr\"odinger picture.
These Markovian QMEs are the direct generalizations of the standard Markovian QMEs for the time-independent Hamiltonian.
Obviously, the obtained QMEs recover the standard QMEs if we assume $\oH_{\S}(t) = \oH_{\S}$.  

\subsection{Validity Beyond Adiabatic Regime}\label{subsec:BeyondAdiabaticRegime}
In the previous section~\ref{subsec:IEA}, we have derived the Markovian QMEs within the IEA.
However, we have to mention that the forms of the dissipators [Eqs.~\eqref{eq:SP::DIEA} and \eqref{eq:SP::DIESA}] have already been known or expected from a long time ago~\cite{Davies78,Childs01,Cai10}.
This is {\em not} surprising because the derived dissipators can be obtained just by intuitively replacing the time-independent variables by the corresponding time-dependent ones.
However, it has been believed that the applicable range is limited to the adiabatic-evolution regime that satisfies the ordinary adiabatic theorem.
Following the recent arguments in Ref.~\onlinecite{Albash12}, for example, the evolution operator $\ou_{\S}(t_1, t_2)$ has been approximated by the `ideal' adiabatic-evolution operator $\ou^{\text{ad}}_{\S}(t_1, t_2)$ [Eq.~\eqref{eq:EVad}] to avoid the difficulty originating from the time convolution between $C_{\alpha\beta}(\tau)$ and $\vS_{\beta}(t -\tau)$.
However, such an approximation is {\em not} needed at all in the formulation shown above.
In this sense, the applicable range is already extended beyond the adiabatic regime.
This is our first important result.

However, based on the formulation above, the IEA is still limited to the time regime of $t$ satisfying $\tau_{\B} \ll \tau_{\A}(t)$.
This means that $\oH_{\S}(t)$ should remain unchanged at least within $\tau_{\B}$ to justify $\oD^{\I\E\A}_t$.
Nevertheless, in a broad range of situations, our claim of the validity is {\em not} restricted to just the regime of $\tau_{\B} \ll \tau_{\A}(t)$ for the IEA; the $\oD^{\I\E\A}_t$ is still well justified even if $\tau_{\B} \ll \tau_{\A}(t)$ fails.

To show this result, we consider the time regime satisfying
\begin{align}
\tau_{\A}(t) \ll \tau_{\R}(t).
\label{eq:NeglectRelaxation}
\end{align}
In this regime, $\oH_{\S}(t)$ is driven much more rapidly than the relaxation time scale.
This means that $\oD^{\I\E\A}_t$ [the second term in Eq.~\eqref{eq:SP::QMEIEA}] has only a minor effect on the dynamics, compared to the Hamiltonian dynamics [the first term in Eq.~\eqref{eq:SP::QMEIEA}].
In such a situation, only a rough evaluation of the superoperator $\oD^{\I\E\A}_t$ would be sufficient.
In other words, the effect of the relaxation is negligible if we focus on the dynamics in the timescale of $\tau_{\A}(t)$ when $\tau_{\A}(t) \ll \tau_{\R}(t)$.
We can therefore state that the QME with $\oD^{\I\E\A}_t$ [Eqs.~\eqref{eq:SP::QMEIEA} and \eqref{eq:SP::DIEA}] strongly breaks down only when the IEA is invalid [$\tau_{\A}(t) \lesssim \tau_{\B}$], and simultaneously, the effect of the relaxation is non-negligible [$\tau_{\R}(t) \lesssim \tau_{\A}(t)$], namely,
\begin{align}
\tau_{\R}(t) \lesssim \tau_{\A}(t) \lesssim \tau_{\B}.
\label{eq:QMEbreak}
\end{align}
However, this condition can never be satisfied as long as the sufficient condition for the WCA holds, as shown below.

For the WCA, which is correct in the second order of $\oH_{\S\B}$, the fourth order contribution must sufficiently be small in comparison to the second order~\cite{McCutcheon10} because the third order contribution vanishes due to the assumption of Eq.~\eqref{eq:OddMoments}.
Here, the second order contribution is roughly estimated by $\tau_{\B,\alpha\beta} C_{\alpha\beta}(0)$ in the $\tau$-integral of Eq.~\eqref{eq:MQME}, whereas the corresponding forth order is similarly estimated by $(\tau_{\B,\alpha\beta})^3 \{C_{\alpha\beta}(0)\}^2$.
The WCA therefore remains valid when $\tau_{\B,\alpha\beta} C_{\alpha\beta}(0) \gg (\tau_{\B,\alpha\beta})^3 \{C_{\alpha\beta}(0)\}^2$, or equivalently,
\begin{align}
(\tau_{\B,\alpha\beta})^2 \textstyle{\int^{\infty}_{-\infty} \frac{\dd \omega}{2\pi}} \gamma_{\alpha\beta}(\omega) \ll 1,
\label{eq:WCA}
\end{align}
for all $\alpha$ and $\beta$.
Here, due to the Fourier-transform relation between $\gamma_{\alpha\beta}(\omega)$ and $C_{\alpha\beta}(\tau)$, the spectral bandwidth of $\gamma_{\alpha\beta}(\omega)$, denoted by $\omega_{\B,\alpha\beta}$, satisfies $\tau_{\B,\alpha\beta} \simeq 2\pi/\omega_{\B,\alpha\beta}$.
We can therefore estimate $\int^{\infty}_{-\infty} \frac{\dd \omega}{2\pi} \gamma_{\alpha\beta}(\omega) \simeq \gamma_{\peak,\alpha\beta}/\tau_{\B,\alpha\beta}$, where $\gamma_{\peak,\alpha\beta}$ is the peak value of $\gamma_{\alpha\beta}(\omega)$.
Putting this into Eq.~\eqref{eq:WCA}, we can find
\begin{align}
\tau_{\B} \ll \tau_{\R}^{\text{min}},
\label{eq:WCAcondition}
\end{align}
as the sufficient condition for the WCA, where $\tau_{\R}^{\text{min}}$ is defined by 
\begin{align}
(\tau_{\R}^{\text{min}})^{-1} \equiv \max_{\alpha,\beta} \gamma_{\peak,\alpha\beta}.
\label{eq:tauRmin}
\end{align}
Since $\tau_{\R}^{\text{min}} \le \tau_{\R}(t)$ holds by definition, the sufficient condition for the WCA [Eq.~\eqref{eq:WCAcondition}] results in 
\begin{align}
\tau_{\B} \ll \tau_{\R}^{\text{min}} \le \tau_{\R}(t),
\end{align}
which is, however, incompatible with Eq.~\eqref{eq:QMEbreak}.
Therefore, if we limit ourselves to the WCA regime, Eq.~\eqref{eq:QMEbreak} can never be satisfied.
Thus, the strong breakdown of the QME with $\oD^{\I\E\A}_t$ [Eqs.~\eqref{eq:SP::QMEIEA} and \eqref{eq:SP::DIEA}] is entirely avoidable as long as the WCA condition is satisfied.
Although it is well-known that simple Hamiltonian evolution is sufficient to characterize the system dynamics for $\tau_\A(t) \ll \tau_\R(t)$, the important point here is that $\tau_\A(t) \ll \tau_\R(t)$ is {\em always} satisfied when $\tau_\A(t) \ll \tau_\B$ because $\tau_\B \ll \tau_\R(t)$ under the WCA.
As a result, we obtain our second important result; 
$\oD^{\I\E\A}_t$ is well justified even though $\tau_{\A}(t)$ is comparable to or even much shorter than $\tau_{\B}$, i.e.~$\tau_{\A}(t) \lesssim \tau_{\B}$.
Since $\oD^{\I\E\A}_t$ is well justified also for $\tau_\B \ll \tau_\A(t)$  [Eq.~\eqref{eq:IEAcondition}] as described in Section~\ref{subsec:IEA}, this means that the Markovian QME can be justified regardless of the speed of $\oH_S(t)$.
This is in contrast to the previous studies \cite{Vega10, Albash12, Davies78, Cai10, Kamleitner13, Pekola10, Vega15, Whitney11, Nalbach14, Javanbakht15} in which it is believed that the Markovian description breaks down when $\oH_S(t)$ changes more rapidly than $\tau_\B$.
The applicable range of the IEA can thus be significantly extended to the regime where the temporal change of $\oH_{\S}(t)$ is faster than $\tau_{\B}$.

Nevertheless, we have to take care the accumulated time of the IEA being invalid, i.e.~$\tau_{\A}(t) \lesssim \tau_{\B}$.
This is because the Markovian QME still weakly breaks down in this regime and the error would be accumulated.
Therefore, the presented approach of $\oD^{\I\E\A}_t$ would not be applicable if the accumulated time becomes comparable to the relaxation time. 
In that case, we should return to Eq.~\eqref{eq:MQME} even though its numerical cost is not low in general.
However, even in such a case, we stress that our discussions on the adiabaticity of $\oH_{\S}(t)$ (Section~\ref{sec:Adiabaticity}) give a clear guide to reduce the numerical effort in the following way.
Since $\tau_{\A}(t)$ is now well defined by  Eqs.~\eqref{eq:tauA}--\eqref{eq:tauAS}, we can calculate $\tau_{\A}(t)$ simultaneously with the density operator $\oRho_{\S}(t)$.
Then, if the IEA is valid at this moment, $\tau_{\B} \ll \tau_{\A}(t)$, the next time step can be obtained, based on $\oD^{\I\E\A}_t$.
If not, $\tau_{\B} \gtrsim \tau_{\A}(t)$, the next time step will be calculated by Eq.~\eqref{eq:MQME}.
Thus, it is only the time region $\tau_{\B} \gtrsim \tau_{\A}(t)$ that requires Eq.~\eqref{eq:MQME} and the numerical cost may be greatly reduced by this approach.

Finally, let us mention the applicability of the SA in a similar line of thought to the above discussion.
To perform the SA, we showed that $\tau_{\S}(t) \ll \tau_{\R}(t)$ and $\tau_{\S}(t) \ll \tau_{\A}(t)$ [Eq.~\eqref{eq:IEA::SA}] are further required.
The first condition is not related to the adiabaticity of $\oH_{\S}(t)$, and therefore, we assume that $\tau_{\S}(t) \ll \tau_{\R}(t)$ is indeed satisfied.
In contrast, the second condition breaks down when $\oH_{\S}(t)$ rapidly varies within $\tau_{\S}(t)$.
However, the dissipator plays only a minor role if $\tau_{\A}(t) \ll \tau_{\R}(t)$ [Eq.~\eqref{eq:NeglectRelaxation}] as discussed above.
Therefore, the QME with $\oD^{\I\E\S\A}_t$ strongly breaks down only when $\tau_{\S}(t) \ll \tau_{\A}(t)$ fails [$\tau_{\A}(t) \lesssim \tau_{\S}(t)$], and simultaneously, the effect of the relaxation is non-negligible [$\tau_{\R}(t) \lesssim \tau_{\A}(t)$], namely,
\begin{align*}
\tau_{\R}(t) \lesssim \tau_{\A}(t) \lesssim  \tau_{\S}(t).
\end{align*}
However, this condition can never be satisfied as long as the first condition, $\tau_{\S}(t) \ll \tau_{\R}(t)$, holds.
As a result, $\oD^{\I\E\S\A}_t$ is still well justified as long as $\tau_{\S}(t) \ll \tau_{\R}(t)$ if the accumulated time of $\tau_{\S}(t) \ll \tau_{\A}(t)$ being invalid is still much shorter than the relaxation timescale.

\section{Application to the Dissipative Landau-Zener Model}\label{sec:DLZ}
To demonstrate the presented general ideas, in the following, we apply the Markovian QME to the DLZ model, in which the Landau-Zener (LZ) transition in a dissipative environment is studied.
This model has been studied in various contexts~\cite{Kayanuma84,Ao89,Wubs06,Saito07,Pokrovsky07,Nalbach09,Whitney11,Nalbach14,Javanbakht15} because it provides the simplest model to describe the adiabatic and non-adiabatic transitions at an avoided level crossing with dissipation.
Especially, in the AQC and the QA, the DLZ model plays a key role in understanding their computational ability.
This is because their practical performance is essentially determined by the non-adiabatic transitions at the minimum gap between the ground state and the first excited state in the presence of noise~\cite{Dickson13, Pudenz14, Amin08}.
Furthermore, the model is appropriate for our purpose because the exact solution is known at $T=0$~\cite{Wubs06,Saito07}.

\begin{figure}[!tb] 
\centering
\includegraphics[width=.49\textwidth]{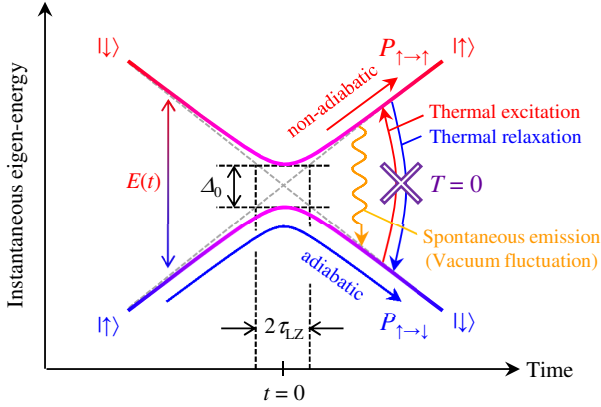}
\caption{(Color online)
The dissipative Landau-Zener model.
The system is initially assumed to be $\ket{\uparrow}$ at $t=-\infty$.
Then, the transition probability $P_{\uparrow \to \downarrow}$ ($P_{\uparrow \to \uparrow}$) to find the system in the ground (excited) state at $t=\infty$ is discussed.
The difference between the two eigenenergies is given by $E(t) = \sqrt{\varDelta^2_0 + (vt)^2}$.
$\tau_{\LZ} \equiv \varDelta_0/v$ denotes the characteristic time for the eigenstates to pass through the minimum gap $\varDelta_0$ around $t = 0$.
When the bath is at zero temperature ($T = 0$), there is no thermal excitation and relaxation but only the spontaneous emission can occur due to the vacuum fluctuation.
}
\label{fig:DLZmodel}
\end{figure}

The system Hamiltonian of the DLZ model is described by 
\begin{align*}
\oH_{\S}(t) = \frac{vt}{2}\oSgm_{z} + \frac{\varDelta_0}{2}\oSgm_{x},
\end{align*}
where $v$ is the LZ sweep velocity, $\varDelta_0 > 0$ is the constant tunneling amplitude, and $\oSgm_{x,z}$ describe the Pauli operators.
We denote the eigenstates of $\oSgm_{z}$ by $\ket{\uparrow/\downarrow}$, i.e. $\oSgm_{z}\ket{\uparrow/\downarrow}=\pm\ket{\uparrow/\downarrow}$.
A schematic illustration of the model is shown in Fig.~\ref{fig:DLZmodel}.
Here, $E(t) = \sqrt{\varDelta_0^2 + (vt)^2}$ denotes the difference between the two eigenenergies and we can see the avoided level crossing at $t = 0$.
The characteristic time for the eigenstates to pass through the minimum gap $\varDelta_0$ around $t = 0$ is described by
\begin{align*}
\tau_{\LZ} \equiv \varDelta_0/v.
\end{align*}
The quantities of our interest is then the adiabatic and non-adiabatic transition probabilities $P_{\uparrow \to \downarrow}$ and $P_{\uparrow \to \uparrow}$ with dissipation by assuming that the system is initially in $\ket{\uparrow}$ at $t = -\infty$. 
Note that $P_{\uparrow \to \downarrow} + P_{\uparrow \to \uparrow} = 1$ by definition.

In this situation, the timescales related to $\oH_{\S}(t)$ (Section~\ref{sec:Adiabaticity}) are analytically obtained as 
\begin{align}
\tau_{\A\S}(t) = \frac{2E^2(t)}{v\varDelta_0}, \quad
\tau_{\A\E}(t) = \frac{E^2(t)}{v^2|t|},
\label{eq:DLZ::tauAS+AE}
\end{align}
and
\begin{align}
&\tau_{\E}(t) = \tau_{\S}(t) =  \frac{1}{E(t)},
\label{eq:DLZ::tauE+S}
\end{align}
according to the definitions listed in Table~\ref{tbl:timescales}.
Then, $\tau_{\A}(t)$ is given by 
\begin{align}
\tau_{\A}(t) = 
\begin{cases}
\tau_{\A\S}(t) & \text{for $|t| < \tau_{\LZ}/2$}  \\
\tau_{\A\E}(t) & \text{for $|t| \ge \tau_{\LZ}/2$} 
\end{cases}.
\label{eq:DLZ::tauA}
\end{align}
Here, we note that $\tau_{\A\S}(t)$ reaches its minimum value $2\tau_{\LZ}$ at $t = 0$, while $\tau_{\A\E}(t)$ reaches the same minimum value $2\tau_{\LZ}$ at $t = \pm \tau_{\LZ}$.
Therefore, the minimum value of $\tau_{\A}(t)$ also becomes $2\tau_{\LZ}$, which is achieved not only at $t=0$ but also at $t = \pm \tau_{\LZ}$ [cf. Fig.~\ref{fig:Evolution}(d)].
We remark that the expression of $\tau_{\A}(t)$ cannot be obtained without  Eqs.~\eqref{eq:tauA}--\eqref{eq:tauAS} even in this simple model and this is why there has been no consensus in the previous studies when discussing the timescale of $\oH_{\S}(t)$.

\begin{figure}[!tb] 
\centering
\includegraphics[width=.49\textwidth]{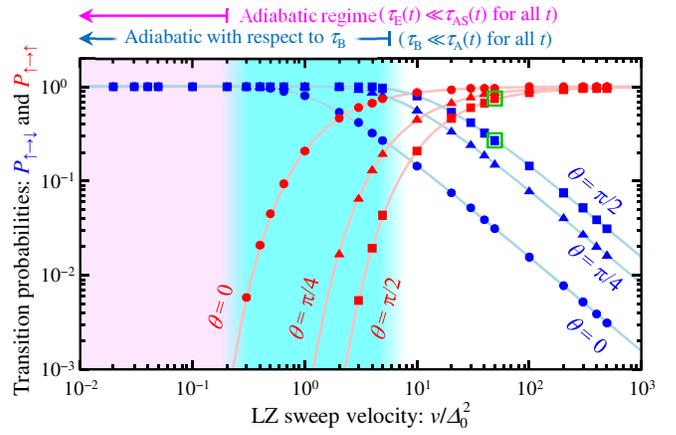}
\caption{(Color online)
$P_{\uparrow \rightarrow \downarrow}$ (blue) and $P_{\uparrow \rightarrow \uparrow}$ (red) at $T=0$, as a function of the LZ sweep velocity $v$.
Symbols: the results of the Markovian QME.
We remark that the imaginary part of $\varGamma(\omega)$ (the Lamb shift) is not neglected in the calculation.
Solid lines: the exact probabilities of $P_{\uparrow \rightarrow \downarrow}=1-P_{\uparrow \rightarrow \uparrow} = 1 - \exp(-\pi W^2/2v)$ with $W^2 = \{ \varDelta_0 - \int^{\infty}_{0} \dd\omega \sin\theta \cos \theta \frac{J(\omega)}{\omega} \}^2 + \int^{\infty}_{0} \dd\omega \sin^2 \theta J(\omega)$~\cite{Wubs06,Saito07}.
Note that $\theta = 0$ (diagonal coupling) gives $W^2 = \varDelta_0^2$, which results in $P_{\uparrow \rightarrow \downarrow} = 1 - \exp(-\pi \varDelta_0^2/2v)$.
This is exactly the same as the LZ transition probability {\em without dissipation}, as pointed out in Ref.~\onlinecite{Wubs06}.
Parameters: $\eta=0.01$ and $\omega_{\c}=30\varDelta_0$.
The left arrows indicate the regimes by taking $\tau_{\X}(t) \ll \tau_{\Y}(t)$ as $\tau_{\X}(t) \lesssim \tau_{\Y}(t)/10$ for all $t$ ($\X \in \{\E, \B\}$, $\Y \in \{\A\S, \A\}$).
One is the ordinary adiabatic-evolution regime and the other is the adiabatic regime with respect to $\tau_{\B}$.
The green open squares correspond to the time evolutions in Fig.~\ref{fig:Evolution}.
}
\label{fig:Probabilities}
\end{figure}

The system is further coupled to the bosonic bath $\oH_{\B}=\sum_{j} \omega_{j} \db_{j}\ob_{j}$ with the interaction Hamiltonian,
\begin{align}
\oH_{\S\B} = \sum_{j}\frac{g_{j}}{2}(\cos \theta \oSgm_z + \sin \theta \oSgm_x)(\ob_j+\db_j),
\label{eq:HSB_DLZ}
\end{align}
where $\theta$ describes the coupling angle and $g_j$ is the system-bath coupling strength.
For definiteness, in our analysis, the spectral density defined by $J(\omega) \equiv \sum_{j}g_j^2\delta(\omega - \omega_j)$ is assumed to be the Ohmic one,
\begin{align*}
J(\omega) = \eta \omega e^{-\omega/\omega_\c},
\end{align*}
with a cutoff energy $\omega_\c$ and a dimensionless coefficient $\eta$.
In comparison with the general form of $\oH_{\S\B}$ [Eq.~\eqref{eq:HSB}], Eq.~\eqref{eq:HSB_DLZ} allows us to define 
\begin{align*}
\oS = \frac{1}{2}(\cos \theta \oSgm_z + \sin \theta \oSgm_x), \quad \oB = \sum_{j}g_{j}(\ob_j+\db_j),
\end{align*}
where we have dropped the subscript `$\alpha$'.
Therefore, by assuming that the bath $\oRho_{\B}$ is in the thermal equilibrium,
$\gamma(\omega) = \int^{\infty}_{-\infty} \dd \tau e^{\ii\omega\tau} C(\tau)$ with $C(\tau) = \TrB{\vB(\tau) \vB(0) \oRho_{\B}}$ [see below Eq.~\eqref{eq:std::tauR}] is given by
\begin{align}
\frac{\gamma(\omega)}{2\pi} = n_{\B}(-\omega)J(-\omega)\varTheta(-\omega) + [n_{\B}(\omega)+1] J(\omega)\varTheta(\omega),
\label{eq:DLZ::gamma}
\end{align}
where $n_{\B}(\omega)=1/(e^{\omega/T}-1)$ is the Bose distribution, $\varTheta(\omega)$ is the step function.
Here, it is important to note that, if the SA is performed, one can find that $\gamma(\omega < 0)$ and $\gamma(\omega > 0)$ describe the excitation and relaxation of the system, respectively, while $\gamma(\omega = 0)$ describes the pure dephasing.
This means that the terms proportional to $n_{\B}(-\omega)$ and  $n_{\B}(\omega)$ denote the thermal excitation and relaxation of the system, respectively.
In contrast, the `$+1$' term corresponds to the spontaneous emission due to the vacuum fluctuation of the bath.
In particular, at $T = 0$, only the spontaneous emission term survives in Eq.~\eqref{eq:DLZ::gamma} and $\gamma(\omega) = 2\pi J(\omega) \varTheta(\omega)$ holds.
This situation is also depicted in Fig.~\ref{fig:DLZmodel}.

\begin{figure*}[!tb] 
\centering
\includegraphics{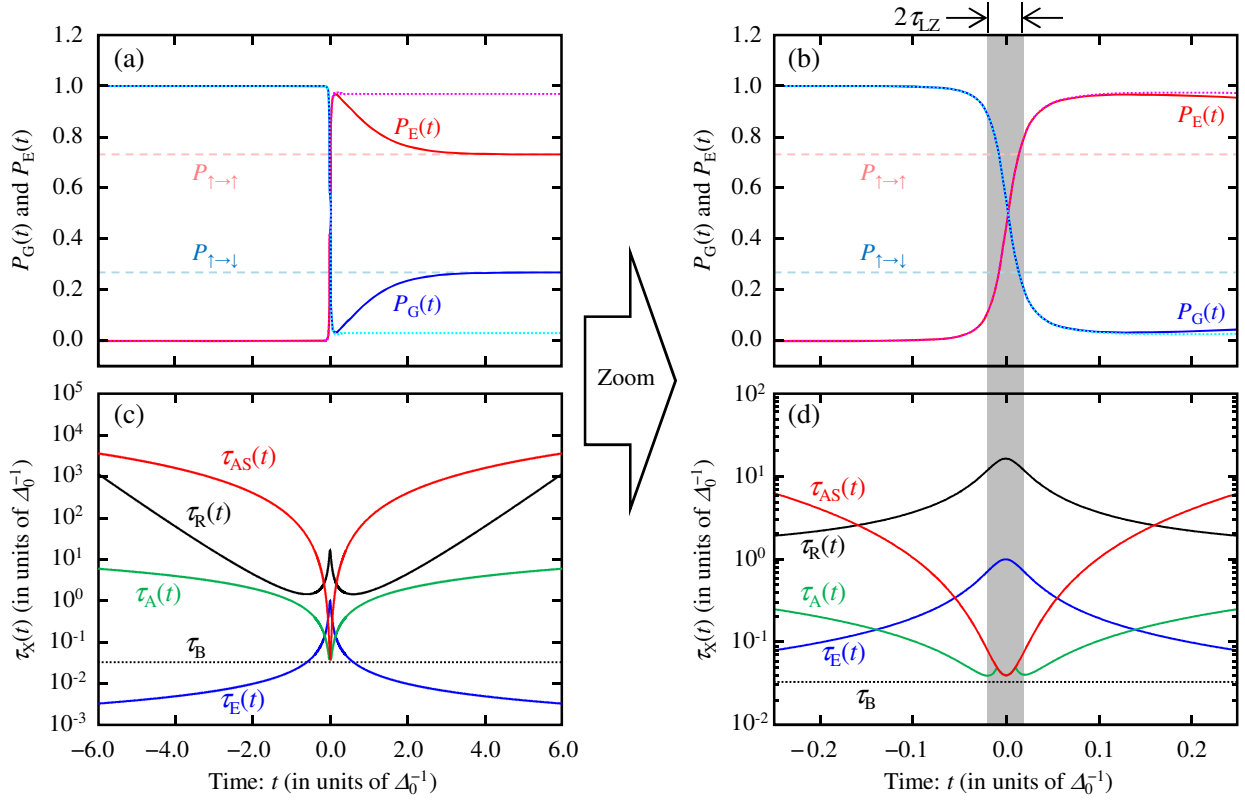}
\caption{(Color online) Time evolutions at $v=50\varDelta_0^2$ for $\theta=\pi/2$.
(a) The probabilities of $P_{\G}(t)$ and $P_{\E}(t)$.
The dotted lines are the corresponding probabilities without dissipation.
The dashed lines show $P_{\uparrow \rightarrow \downarrow}$ and $P_{\uparrow \rightarrow \uparrow}$ for comparison.
Panel~(b) shows a zoom around $t=0$.
(c) The timescales of $\tau_{\A\S}(t)$, $\tau_{\E}(t)$, $\tau_{\A}(t)$, $\tau_{\B}$ and $\tau_{\R}(t)$ in units of $\varDelta_0^{-1}$, according to Eqs.~\eqref{eq:DLZ::tauAS+AE}--\eqref{eq:DLZ::tauA}, Eq.~\eqref{eq:DLZ::tauB} and Eq.~\eqref{eq:IEA::tauR}.
Panel~(d) again shows a zoom around $t=0$.
$\tau_{\A\S}(t)$ reaches its minimum value $2\tau_{\LZ}$ at $t=0$; $\tau_{\A\S}(0) = 2\tau_{\LZ}$.
The minimum value of $\tau_{\A}(t)$ is also $2\tau_{\LZ}$ but achieved not only at $t = 0$ but also $t = \pm \tau_{\LZ}$.
We remark that Panels (c) and (d) should be discussed with reference to Table~\ref{tbl:conditions}.
The parameters are the same as indicated by the green open squares in Fig.~\ref{fig:Probabilities}.
The gray shaded area corresponds to $|t| \leq \tau_{\LZ}$.
}
\label{fig:Evolution}
\end{figure*}

Since the dissipative nature of the system is essentially determined by $J(\omega)$, let us briefly discuss the two parameters $\eta$ and $\omega_\c$ before studying numerical results.
In the case of zero temperature, $\gamma(\omega) = 2\pi J(\omega) \varTheta(\omega)$ gives the bath correlation function $C(\tau)$ by
\begin{align}
C(\tau)=\eta\omega_{\c}^2/(1+\ii\tau\omega_{\c})^2,
\label{eq:DLZ::C(tau)}
\end{align}
where the Fourier transform relation between $\gamma(\omega)$ and $C(\tau)$ has been used.
Hence, we can estimate the decay time of the bath correlation function $C(\tau)$ as
\begin{align}
\tau_{\B} \simeq \omega_{\c}^{-1}.
\label{eq:DLZ::tauB}
\end{align}
We note that this is equivalent to the statement that the spectral band width of $\gamma(\omega)$ can be estimated by $\omega_{\c}$.
As a result, if we set $\omega_{\c} \ll \varDelta_0$, there is no spectral density at the transition energy $E(t)$, i.e. $\gamma(E(t)) = 2\pi J(E(t)) \simeq 0$, and the dissipative nature plays no significant role. 
For our purpose, therefore, we should set $\omega_{\c} \gtrsim \varDelta_0$.
Moreover, by using Eqs.~\eqref{eq:DLZ::C(tau)} and \eqref{eq:DLZ::tauB}, we can estimate the condition for the WCA,
\begin{align}
\eta \ll 1,
\label{eq:DLZ::WCA}
\end{align}
due to Eq.~\eqref{eq:WCA}.
The WCA becomes valid when this condition is satisfied.
Based on these considerations, we have performed numerical calculations by setting $\eta = 0.01$ and $\omega_{\c} = 30\varDelta_0$.
Although the SA will not be applied in our calculation, we note that the SA ($\oD^{\I\E\A}_t \simeq \oD^{\I\E\S\A}_t$) does not show any significant difference in the parameter range presented below.

Figure~\ref{fig:Probabilities} shows the transition probabilities $P_{\uparrow \rightarrow \downarrow}$ and $P_{\uparrow \rightarrow \uparrow}$ as a function of the LZ sweep velocity $v$.
For comparison, the exact results are also shown by solid lines, where the transition probability for $\theta = 0$ (diagonal coupling) is exactly the same as the LZ transition probability {\em without dissipation}~\cite{Wubs06}.
In Fig.~\ref{fig:Probabilities}, when $\tau_{\E}(t) \ll \tau_{\A\S}(t)$ for all $t$ (the pink shaded area), the LZ sweep velocity $v$ is slow enough to satisfy the adiabatic theorem.
In this regime, the spontaneous emission does not play any role because the state is always in the instantaneous ground state.
As a result, $P_{\uparrow \to \downarrow} \simeq 1$ holds for all $\theta$. 
By increasing the velocity $v$, the condition for the adiabatic theorem is violated and the non-adiabatic transition becomes discernible.
In this regime that still satisfies $\tau_{\B} \ll \tau_{\A}(t)$ for all $t$ (the aqua shaded region), $P_{\uparrow \to \downarrow}$ decreases from one, and instead, $P_{\uparrow \to \uparrow}$ increases.
Here, we can notice that $P_{\uparrow \to \downarrow}$ for $\theta = \pi/2$ and $\pi/4$ are greater than for $\theta = 0$.
This means that, for $\theta = \pi/2$ and $\pi/4$, the ground state is recovered by the spontaneous emission even though the non-adiabatic transition is possible.
By increasing the velocity $v$ further ($v/\varDelta_0^2 \gtrsim 6.0$), the decrease of $P_{\uparrow \to \downarrow}$ becomes pronounced for all $\theta$ and $P_{\uparrow \to \uparrow}$ approaches one.
In this regime, the recovery of the ground state becomes incomplete because the transition energy $E(t)$ rapidly goes through the spectral bandwidth of $\gamma(\omega)$ before the ground state is sufficiently recovered.

Now, it is important to point out that excellent agreements between the exact probabilities (solid lines) and the results of the Markovian QME (symbols) are obtained, regardless of the speed of $\oH_\S(t)$.
This is notable in the following two points.
First, it has been believed that the Markovian description cannot capture the exact probabilities of the DLZ model due to the non-Markovian effect~\cite{Kamleitner13,Javanbakht15}.
Nevertheless, Fig.~\ref{fig:Probabilities} clearly shows that the Markovian description is sufficient at least in the weak coupling regime.
Second, more importantly, the agreements are remarkable even if the temporal change of $\oH_{\S}(t)$, i.e.~$\tau_{\A}(t)$, is comparable to or even faster than $\tau_{\B}$.
This is in stark contrast to the conventional expectation that the Markovian QME is applicable only when the system Hamiltonian varies much more slowly than $\tau_{\B}$.
Figure~\ref{fig:Probabilities} clearly shows that the Markovian QME still works well not only beyond the ordinary adiabatic-evolution regime but also beyond the slowly-varying regime with respect to $\tau_{\B}$, which supports the validity of our general framework.
Here, we note that the exact solutions (for $\theta \ne 0$) cannot be reproduced if the simple Hamiltonian evolution is just applied instead of the Markovian QME although one may expect that there is no need of the Markovian QME in such a rapid regime.

To further shed light on the underlying mechanism, in Figs.~\ref{fig:Evolution}(a) and \ref{fig:Evolution}(b), we show the time-dependent probabilities $P_{\G}(t)$ and $P_{\E}(t)$ to find the system in the instantaneous ground and excited states, respectively.
In the calculation, we set $v = 50\varDelta_0^2$ and $\theta = \pi/2$, the situation of which corresponds to the green open squares in the rapid regime of Fig.~\ref{fig:Probabilities}.
For $t \lesssim 0$ in Fig.~\ref{fig:Evolution}(a), $P_{\G}(t) \simeq 1$ and $P_{\E}(t) \simeq 0$ hold true and the values are almost the same as the corresponding probabilities without dissipation (the dotted lines).
In this regime, therefore, neither the non-adiabatic transition nor the dissipation play significant roles.
In the vicinity of $t \simeq 0$, $P_{\G}(t)$ changes from one toward zero and $P_{\E}(t)$ from zero toward one.
The zoom around $t \simeq 0$ [Fig.~\ref{fig:Evolution}(b)] shows that this non-adiabatic transition occurs mainly in the regime $|t| \lesssim \tau_{\LZ}$ (the gray shaded area).
Here, one finds that the non-adiabatic transition is still close to the behavior without dissipation.
This suggests that the non-adiabatic transition is essentially instantaneous with respect to the timescale of relaxation.
Finally, for $t \gtrsim 0$ in Fig.~\ref{fig:Evolution}(a), $P_{\G/\E}(t)$ gradually approaches the value of $P_{\uparrow \rightarrow \downarrow/\uparrow}$ due to the dissipation.
However, in this time period, there is no non-adiabatic transition, as evidenced by the fact that the probabilities without dissipation (the dotted lines) do not show any change in time.
As a result, we can notice that there is no time where the non-adiabatic transition and the dissipation simultaneously play significant role.

This situation can be studied more carefully and clearly from the viewpoint of the individual timescales summarized in Table~\ref{tbl:conditions}.
In Figs.~\ref{fig:Evolution}(c) and \ref{fig:Evolution}(d), we therefore show $\tau_{\X}(t)$ with $\X \in \{ \A\S, \E, \A, \B, \R \}$, according to Eqs.~\eqref{eq:DLZ::tauAS+AE}--\eqref{eq:DLZ::tauA}, Eq.~\eqref{eq:DLZ::tauB} and Eq.~\eqref{eq:IEA::tauR}.
For $t \lesssim 0$ in Fig.~\ref{fig:Evolution}(c), we can find $\tau_{\E}(t) \ll \tau_{\A\S}(t)$.
This indicates that the ordinary adiabatic evolution is indeed validated, which is consistent with the above discussion.
In addition, we also find that $\tau_{\B} \ll \tau_{\A}(t)$ holds true for $t \lesssim 0$.
Therefore, according to Table~\ref{tbl:conditions}, we notice that the IEA is also well justified in this time regime.

However, around $t \simeq 0$, one finds that these conditions break down.
As shown in Fig.~\ref{fig:Evolution}(d), the adiabatic evolution is no longer safely ensured for $|t| \lesssim 0.1 \varDelta^{-1}_0$ because $\tau_{\A\S}(t)$ falls within one order of magnitude of $\tau_{\E}(t)$ and $\tau_{\E}(t) \ll \tau_{\A\S}(t)$ is not satisfied. 
In particular, $\tau_{\A\S}(t)$ reaches its minimum value of $2\tau_{\LZ}$ at $t = 0$, which means that the instantaneous eigenstates change most rapidly at $t = 0$ and its timescale is given by $2\tau_{\LZ}$.
Naturally, the non-adiabatic transition becomes pronounced around $|t| \lesssim \tau_{\LZ}$, as we have already seen in Fig.~\ref{fig:Probabilities}(b).

In a similar manner, for the validity of the IEA, we can notice that $\tau_{\A}(t)$ is within one order of magnitude of $\tau_{\B}$ over the entire region in Fig.~\ref{fig:Evolution}(d).
Hence, the IEA is not ensured in this time regime ($|t| \lesssim 0.25\varDelta_0^{-1}$) due to the failure of $\tau_{\B} \ll \tau_{\A}(t)$.
Nevertheless, one can also find that $\tau_{\A}(t)$ is much shorter than $\tau_{\R}(t)$, i.e.~$\tau_{\A}(t) \ll \tau_{\R}(t)$, in Fig.~\ref{fig:Evolution}(d).
This means that the effect of relaxation is negligible compared to the non-adiabatic effect.
In particular, for $|t| \lesssim \tau_{\LZ}$, $\tau_{\A}(t)$ is two orders of magnitude shorter than $\tau_{\R}(t)$ even though $\tau_{\A}(t)$ becomes comparable to $\tau_{\B}$.
As a result, the Markovian QME still works well because $\oD^{\I\E\A}_t$ plays only a minor role around $t \simeq 0$.
This is actually what we have explained in the general discussion (Section~\ref{subsec:BeyondAdiabaticRegime}).
The important point is that such a situation is generally guaranteed as long as we are in the weak coupling regime; the DLZ model corresponds to one prominent example.

Thus, the DLZ model has been discussed to demonstrate the ability of the Markovian QME in this section.
Finally, it would be worth digressing from the main subject to point out that the spontaneous emission (the vacuum fluctuation) has a great importance in the context of the AQC and the QA.
In this case, the two states in the DLZ model are regarded as the ground and the first excited states.
Then, as illustrated in Figs.~\ref{fig:DLZmodel} and \ref{fig:Probabilities}, the spontaneous emission has the ability to recover the ground state when $\theta \ne 0$.
This means that the computational errors caused by the depopulation of the ground state will be naturally corrected.
In other words, a self-healing mechanism is given to the system by the bath.
For finite temperature, according to Eq.~\eqref{eq:DLZ::gamma}, this mechanism can dominate when $T \ll E(t) \lesssim \omega_{\c}$, whereas the thermal excitation and relaxation will activate when $E(t) \lesssim T \lesssim \omega_{\c}$.
In this sense, the mechanism is different from the thermally-assisted QA in Ref.~\onlinecite{Dickson13}, the situation of which is focused on the latter case.

In order to effectively utilize this mechanism, however, the coupling angle $\theta$ in $\oH_{\S\B}$ [Eq.~\eqref{eq:HSB_DLZ}] should be $\theta = \pi/2$ (transverse coupling) in principle for the following two reasons.
First, the thermal excitation is reduced around the minimum gap because $[\oH_{\S}(t), \oH_{\S\B}] = 0$ at $t=0$.
Second, if $T \ll E(t) \lesssim \omega_{\c}$ is fulfilled, the spontaneous emission can dominate the relaxation process after the avoided crossing because $[\oH_{\S}(t), \oH_{\S\B}] \ne 0$ for $t \gtrsim \tau_{\LZ}$.
In reality, however, it is difficult to directly control $\theta$ (the form of $\oH_{\S\B}$) in experiments.
Therefore, the past discussions were focused only on the diagonal coupling in most cases~\cite{Amin08, Albash12}.
Nevertheless, the encoding direction of the target Hamiltonian of the system would be allowed to be changed, which is physically equivalent to the change of the coupling angle $\theta$.
In this context, we conjecture that the encoding direction also plays an important role to increase the computational performance of the AQC and the QA.

\section{Conclusions and outlook}\label{sec:Conclusions}
We have presented a detailed analysis of the Markovian QME under the WCA when the system Hamiltonian is time-dependent. 
While this problem has been discussed by many authors in the past, there has been no consensus for the condition to validate the formalism.
This fact indicates that the problem is non-trivial, while one may think it straightforward at first glance.
In our view, one major reason was the complete lack of the way to explicitly quantify the temporal change of $\oH_{\S}(t)$.
Therefore, in this paper, we first introduced the timescale $\tau_{\A}(t)$ as a measure for the adiabaticity of $\oH_{\S}(t)$ itself.
Here, the adiabaticity of $\oH_{\S}(t)$ is conceptually different from the ordinary adiabatic theorem, and therefore, enables us to derive the Markovian QME without the adiabatic theorem.
Furthermore, in a broad range of situations, it was also shown that the framework is well justified even if $\tau_{\A}(t)$ is much shorter than the decay timescale of the bath correlation function $\tau_{\B}$.
This result arises from the fact that there is no situation where the non-adiabatic effect and the dissipative nature play considerable roles simultaneously in time, as long as the WCA is validated.
We have thus clearly shown that there is no need to restrict ourselves to either the adiabatic-evolution regime or the slowly-varying regime with respect to $\tau_{\B}$.
This is in stark contrast to the past understanding.
As a result, the Markovian QME is justified well beyond the adiabatic regime.
We here remark that the presented route of the formulation and the well-defined approximations allow us to clearly understand the structure of the framework with sufficient generality.
Hence, the scheme in this paper is immediately applicable to a wide range of physical systems.

As an example, we have applied the framework to the DLZ model and illustrated the ability of the Markovian QME.
Even in this simple model, we stressed that the expression of $\tau_{\A}(t)$ is difficult to obtain without Eqs.~\eqref{eq:tauA}--\eqref{eq:tauAS} and this has been the very origin of the inconsistent quantification of the timescale of $\oH_{\S}(t)$.
Then, it was shown that the numerical results have good agreements with the exact transition probabilities.
Furthermore, the relationship between the relevant timescales [Figs.~\ref{fig:Evolution}(c) and \ref{fig:Evolution}(d)] indeed supports our general scenario.
It would be worth noting that such an approach generally enables an easy estimation of whether the individual approximations are justified or not at a certain time $t$.
Finally, a short digression has been made to discuss the importance of the spontaneous emission in the context of the AQC and the QA.
We discussed the possibility that the spontaneous emission provides the self-healing mechanism that naturally corrects the computational errors.

The results presented in this paper would be of interest to those who try to control quantum systems as a function of time~\cite{Farhi01, Amin09-1, Amin09-2, Kadowaki98, Dickson13, Pudenz14, Amin08, Childs01, Vega10, Albash12, Alicki79, Kosloff02, Uzdin15, Chen11, Bason12, Petta10, Petersson10, Cao13} because in reality the quantum systems cannot be free from uncontrolled interactions with environmental degrees of freedom (bath).
Although our demonstration in the DLZ model was focused on the zero temperature limit, the framework is, of course, applicable to finite temperature cases.
Interesting directions for future research are investigations of the non-adiabatic regime of the AQC and the QA in the presence of dissipation.
As conjectured from the DLZ model, there are indeed possibilities to improve the computational performance.
The problem is, however, still non-trivial when the system size and the number of the relevant eigenstates become large.
In that case, we expect that our methodology will play a key role for the development of new schemes beyond the adiabatic regime.

\begin{acknowledgments}
We thank Y. Yamamoto, N. Nagaosa, K. Kamide, Y. Yamada and M. Bamba for fruitful discussions and comments.
This work was funded by ImPACT Program of Council for Science, Technology and Innovation (Cabinet Office, Government of Japan) and by JSPS KAKENHI (Grants No.~26287087).
\end{acknowledgments}

\appendix
\section{Time-convolutionless formalism \label{TCL}}
Here, in order to make the paper self-contained, we describe the time-convolutionless (TCL) framework and show a brief derivation of Eq.~\eqref{eq:nMRedfield1} within the WCA.
We note that the derivation is the same as Ref.~\onlinecite{Breuer02} even though the system Hamiltonian $\oH_{\S}(t)$ is time-dependent.
This section also plays a preliminary role for the development of the slippage technique in Appendix~\ref{SLP}.

For this purpose, we first rewrite the total Hamiltonian as $\oH(t) = \oH_0(t) + \lambda \oH_{\S\B}$, where $\lambda$ is a dimensionless parameter introduced only to easily measure the order of $\oH_{\S\B}$.
In the interaction picture, the total density operator $\vRho(t)$ evolves according to
\begin{align}
\tfrac{\dd}{\dd t} \vRho(t) = -\ii \lambda [\vH_{\S\B}(t), \vRho(t)] \equiv \lambda\vL_{\S\B}(t)\vRho(t),
\label{app:Sch.eq.}
\end{align}
where $\vL_{\S\B}(t)$ denotes the Liouville superoperator in the interaction picture.
For notational convenience, we will write $\vH_{\S\B;\lambda}(t) \equiv \lambda \vH_{\S\B}(t)$ and $\vL_{\S\B;\lambda}(t) \equiv \lambda \vL_{\S\B}(t)$ in the following.
However, we will finally replace  $\vH_{\S\B;\lambda}(t) \to \vH_{\S\B}(t)$ and $\vL_{\S\B;\lambda}(t) \to \vL_{\S\B}(t)$ after the formulation is completed, in order to eliminate the parameter $\lambda$.

\subsection{The Liouville superoperators}
Before discussing the framework in detail, we summarize here the Liouville superoperators.
We denote the Liouville superoperator in the Schr\"odinger picture by
\begin{align}
\oL_{\alpha}(t)\oX \equiv -\ii[\oH_{\alpha}(t) ,\oX],
\label{app:Liouville_Sch}
\end{align}
where $\oX$ is an arbitrary operator.
We note that there is no time dependence in the Hamiltonian $\oH_{\alpha}(t)$ when $\alpha \in \{ \B, \S\B, \S\B;\lambda \}$ but $\oH_{\alpha}(t)$ is described in the time-dependent way because there is no confusion.
The Liouville superoperator satisfies a relation $\dL_{\alpha}(t) = -\oL_{\alpha}(t)$ in the Liouville space, where the adjoint superoperator $\mathcal{A}^{\dagger}$ for any superoperator $\mathcal{A}$ is defined in such a way that $\Tr{(\mathcal{A}^{\dagger}\oX_1)^{\dagger}\oX_2} = \Tr{\oX_1^{\dagger} \mathcal{A} \oX_2}$ holds for arbitrary operators $\oX_1$ and $\oX_2$.
It is then natural to introduce the evolution superoperator by
\begin{align*}
\oU_{\alpha}(t_2, t_1) \equiv \left \{
	\begin{array}{ll}
	\oT_{+} \exp \left\{ + \int^{t_2}_{t_1} \dd s \oL_{\alpha}(s) \right\} & \quad t_2 \geq t_1 \\
	\oT_{-} \exp \left\{ - \int^{t_1}_{t_2} \dd s \oL_{\alpha}(s) \right\} & \quad t_1 > t_2 \\
	\end{array}
\right.,
\end{align*}
where $\oT_{+(-)}$ denotes the chronological (anti-chronological) time ordering for the superoperators. 
The interaction picture of an arbitrary operator is, for example, easily described by using $\oU_{\alpha}(t_2, t_1)$ as
\begin{align}
\vO(t) = \du_0(t,t_0)\oO(t)\ou_0(t,t_0) = \dU_0(t,t_0)\oO(t).
\label{app:IntPicture}
\end{align}
By further introducing the interaction picture of the Liouville superoperator $\vL_{\alpha}(t) \equiv \dU_0(t,t_0)\oL_{\alpha}(t)\oU_0(t,t_0)$, 
we can obtain
\begin{align}
\vL_{\alpha}(t)\oX = -\ii[\vH_{\alpha}(t), \oX],
\label{app:Liouville_Int}
\end{align}
which is consistent with the definition of Eq.~\eqref{app:Sch.eq.}.
In the similar manner to $\oU_{\alpha}(t_2, t_1)$, we therefore define the evolution operator for the interaction picture by
\begin{align*}
\vU_{\alpha}(t_2, t_1) \equiv \left \{
	\begin{array}{ll}
	\oT_{+} \exp \left\{ + \int^{t_2}_{t_1} \dd s \vL_{\alpha}(s) \right\} & \quad t_2 \geq t_1 \\
	\oT_{-} \exp \left\{ - \int^{t_1}_{t_2} \dd s \vL_{\alpha}(s) \right\} & \quad t_1 > t_2 \\
	\end{array}
\right..
\end{align*}
Note that $\vRho(t)$ can be written as $\vRho(t)=\vU_{\S\B;\lambda}(t,t_0)\vRho(t_0)$ when $\vRho(t)$ evolves according to Eq.~\eqref{app:Sch.eq.}.

\subsection{The TCL form of the QME}
We now derive the TCL form of the QME. 
To this end, we first decompose Eq.~\eqref{app:Sch.eq.} as
\begin{align}
\tfrac{\dd}{\dd t}\oP\vRho(t) &= \oP \vL_{\S\B;\lambda}(t)(\oP + \oQ)\vRho(t), 
\label{app:P} \\
\tfrac{\dd}{\dd t}\oQ\vRho(t) &= \oQ \vL_{\S\B;\lambda}(t)(\oP + \oQ)\vRho(t),
\label{app:Q}
\end{align}
where $\oP$ and $\oQ$ are the projection superoperators defined by $\oP \oX \equiv \TrB{\oX} \otimes \oRho_{\B}$ and  $\oQ \equiv 1-\oP$.
The formal solution of Eq.~\eqref{app:Q} can then be described as
\begin{align}
\oQ \vRho(t) =& \vG_{+}(tt_0) \oQ\vRho(t_0) \nonumber\\
&+ \textstyle{\int^{t}_{t_0} \dd t'} \vG_{+}(tt')\oQ\vL_{\S\B;\lambda}(t')\oP\vRho(t'),
\label{app:FormalQ}
\end{align}
where $\vG_{+}(tt') \equiv \oT_{+} \exp \{ \int^{t}_{t'} \dd s \oQ \vL_{\S\B;\lambda}(s) \}$.
By writing $\vRho(t') = \vU_{\S\B;\lambda}(t',t)\vRho(t) = \vU_{\S\B;\lambda}(t',t)(\oP+\oQ)\vRho(t)$, 
we can readily obtain $\oQ \vRho(t) = \vG_{+}(tt_0) \oQ\vRho(t_0) + \vW(t)(\oP+\oQ)\vRho(t)$ with
$\vW(t) \equiv \int^{t}_{t_0} \dd t' \vG_{+}(tt')\oQ\vL_{\S\B;\lambda}(t')\oP \vU_{\S\B;\lambda}(t',t)$.
Then, we obtain
\begin{align}
\oQ \vRho(t) = \tfrac{1}{1-\vW(t)}\vG_{+}(tt_0) \oQ\vRho(t_0) + \left\{ \tfrac{1}{1-\vW(t)}-1 \right\} \oP\vRho(t).
\label{app:Qtemp}
\end{align}
Here, we have assumed the existence of $\{1-\vW(t)\}^{-1}$ because $\vW(t_0)=0$ and $\vW(t)|_{\alpha=0}=0$ suggest that $1-\vW(t)$ may be inverted for sufficiently small couplings and $t-t_0$~\cite{Breuer02}.
Substitution of Eq.~\eqref{app:Qtemp} into Eq.~\eqref{app:P} finally yields the TCL form of the QME,
\begin{align}
\tfrac{\dd}{\dd t} \vRho_{\S}(t) = \TrB{\vK(t) \oP \vRho(t)} + \TrB{\vI(t) \oQ \vRho(t_0)},
\label{app:TCL}
\end{align}
with definitions of the superoperators
\begin{align}
\vK(t) & \equiv \oP \vL_{\S\B;\lambda}(t) \tfrac{1}{1-\vW(t)} \oP,
\label{app:K}\\
\vI(t)  & \equiv \oP \vL_{\S\B;\lambda}(t) \tfrac{1}{1-\vW(t)} \vG_{+}(tt_0),
\label{app:I}
\end{align}
where we have used $\TrB{\oP \vRho(t)} = \vRho_{\S}(t)$.
Eqs.~\eqref{app:TCL}-\eqref{app:I} are exact and local in time although the superoperators $\vK(t)$ and $\vI(t)$ are complicated in general.
However, by assuming the separable initial state $\vRho(t_0)=\oRho(t_0)=\oRho_{\S}(t_0) \otimes \oRho_{\B}$, the second term in Eq.~\eqref{app:TCL} becomes zero because of $\oQ\vRho(t_0)=0$.
In the next subsection, therefore, $\vK(t)$, called the TCL generator, will be estimated up to the second order in $\lambda$, which corresponds to the WCA.

\subsection{The TCL generator within the WCA}
Here, the TCL generator $\vK(t)$ will be expanded in the power series $\vK(t) = \sum_{n=1}^{\infty}\lambda^n\vK_n(t)$ and truncated to the second order in $\lambda$.
For this purpose, $\{1-\vW(t)\}^{-1}$ in Eq.~\eqref{app:K} is expanded as $\{1-\vW(t)\}^{-1} = \sum_{n=0}^{\infty} [\vW(t)]^n$.
Then, we obtain 
\begin{align*}
\vK(t) = \lambda \sum_{n=0}^{\infty}  \oP \vL_{\S\B}(t) [\vW(t)]^n \oP.
\end{align*}
Therefore, further expansion of $\vW(t)$ in the power series of $\vW(t) = \sum_{n=1}^{\infty}\lambda^n\vW_n(t)$ gives
\begin{align*}
\vK(t) = \lambda \vK_1(t) + \lambda^2 \vK_2(t) + \mathcal{O}(\lambda^3),
\end{align*}
with 
\begin{align*}
\vK_1(t) &= \oP\vL_{\S\B}(t)\oP,\\
\vK_2(t) &=  \textstyle{ \int_{t_0}^{t} \dd t'} \oP\vL_{\S\B}(t) \oQ\vL_{\S\B}(t') \oP.
\end{align*}
As a result, by assuming the vanishing odd moments of the system-bath interaction Hamiltonian $\TrB{ \vH_{\S\B}(t_1) \vH_{\S\B}(t_2) \cdots \vH_{\S\B}(t_{2n+1}) \oRho_{\B} } = 0$ [Eq.~\eqref{eq:OddMoments}], or equivalently, $\oP \vL_{\S\B}(t_1) \vL_{\S\B}(t_2) \cdots \vL_{\S\B}(t_{2n+1}) \oP = 0$, we can find
\begin{align}
\vK(t) = \textstyle{ \int_{t_0}^{t} \dd t'} \oP\vL_{\S\B;\lambda}(t) \vL_{\S\B;\lambda}(t') \oP + \mathcal{O}(\lambda^3).
\label{app:KwithWCA}
\end{align}
The TCL form of the QME [Eq.~\eqref{app:TCL}] then becomes 
\begin{align*}
\tfrac{\dd}{\dd t} \vRho_{\S}(t) &= - \textstyle{ \int_{t_0}^{t} \dd t'} \TrB{ \vH_{\S\B;\lambda}(t), [\vH_{\S\B;\lambda}(t'), \vRho_{\S}(t) \otimes \oRho_{\B}] }.
\end{align*}
Thus, up to the second order in $\oH_{\S\B}$, namely within the WCA, Eq.~\eqref{eq:nMRedfield1} has been derived.

\section{Slippage technique \label{SLP}}
The concept of the slippage technique is first proposed by Su\'arez {\it et al}.~\cite{Suarez92} and expanded by Gaspard and Nagaoka~\cite{Gaspard99} in order to cure the problem of the complete positivity when the secular approximation is not applied.
The slippage of the initial condition alters the original initial condition of the system $\oRho_{\S}(t_0)$ to the slipped (renormalized) one $\SLP\oRho_{\S}(t_0)$, where $\SLP$ is the slippage superoperator.

In this Appendix~\ref{SLP}, the concept is generalized to the case of the time-dependent system Hamiltonian and is employed to justify the Markov approximation.
We note that the slipped initial condition has been used to obtain all of the numerical results presented in Section \ref{sec:DLZ}.

\subsection{The correlation superoperator}
To study the slippage superoperator $\SLP$, by using Eq.~\eqref{app:KwithWCA}, we rewrite the first term of Eq.~\eqref{app:TCL} as
\begin{align*}
\TrB{\vK(t) \oP \vRho(t)} \simeq \textstyle {\int_{t_0}^{t} \dd t'} \dU_{\S}(t,t_0) \oC_{\S}(t, t') \oU_{\S}(t,t_0) \vRho_{\S}(t),
\end{align*}
where $\oC_{\S}(t_2, t_1)$ is the correlation superoperator defined as
\begin{align}
\oC_{\S}(t_2,t_1) \equiv \TrB{\oL_{\S\B} \dU_{0}(t_1,t_2) \oL_{\S\B} \oU_{0}(t_1,t_2)\oRho_{\B}}.
\end{align}
In the derivation, we have used
\begin{align*}
\vL_{\S\B}(t)\vL_{\S\B}(t') = \dU_0(t,t_0) \oL_{\S\B} \dU_0(t',t)\oL_{\S\B} \oU_0(t',t)\oU_0(t, t_0).
\end{align*}
We can therefore rewrite the TCL form of the QME [Eq.~\eqref{app:TCL}] within the WCA as
\begin{align}
\tfrac{\dd}{\dd t} \vRho_{\S}(t) = \textstyle {\int_{0}^{\varDelta t} \dd \tau} \dU_{\S}(t, t_0) \oC_{\S}(t,t-\tau) \oU_{\S}(t,t_0) \vRho_{\S}(t),
\label{app:TCLwithinWCA}
\end{align}
where $\varDelta t = t - t_0$.
This equation should be compared with Eq.~\eqref{eq:nMRedfield2}.
One can then readily notice the following relation;
\begin{multline}
\dU_{\S}(t,t_0) \oC_{\S}(t,t-\tau) \oU_{\S}(t,t_0)\vRho_{\S}(t) \\
= \textstyle{\sum_{\alpha, \beta} } C_{\alpha\beta}(\tau)
 \{ \vS_{\beta}(t-\tau) \vRho_{\S}(t) \dvS_{\alpha}(t) \\
- \dvS_{\alpha}(t)\vS_{\beta}(t-\tau) \vRho_{\S}(t)\} + \Hc,
\label{app:temp1}
\end{multline}
which will be used to derive the slippage superoperator in the next subsection.
Here, under the Markovian approximation ($\varDelta t \to \infty$), Eq.~\eqref{app:TCLwithinWCA} gives
\begin{align}
\tfrac{\dd}{\dd t} \vRho_{\S}^{\M}(t) = \textstyle {\int_{0}^{\infty} \dd \tau} \dU_{\S}(t, t_0) \oC_{\S}(t,t-\tau) \oU_{\S}(t, t_0) \vRho_{\S}^{\M}(t),
\label{app:TCLwithinWCA_MA}
\end{align}
where $\vRho_{\S}^{\M}(t)$ is distinguished from $\vRho_{\S}(t)$ because $\vRho_{\S}^{\M}(t)$ may be different from $\vRho_{\S}(t)$ especially in the time region of $\varDelta t \lesssim \tau_{\B}$.

\subsection{The slippage superoperator}
We are now ready to derive the slippage superoperator $\SLP$.
To this end, we formally integrate Eq.~\eqref{app:TCLwithinWCA} and Eq.~\eqref{app:TCLwithinWCA_MA} in time, respectively,
\begin{align}
\vRho_{\S}(t) & = \oRho_{\S}(t_0) \nonumber\\
 + &\textstyle {\int_{t_0}^{t} \dd t'   \int_{0}^{t'-t_0} \dd \tau} \dU_{\S}(t',t_0) \oC_{\S}(t',t'-\tau) \oU_{\S}(t',t_0) \vRho_{\S}(t'),
\label{app:nonMarkov}\\
\vRho_{\S}^{\M}(t) & = \oRho_{\S}^{\M}(t_0) \nonumber\\
+ &\textstyle {\int_{t_0}^{t} \dd t'   \int_{0}^{\infty} \dd \tau} \dU_{\S}(t',t_0) \oC_{\S}(t',t'-\tau) \oU_{\S}(t',t_0) \vRho_{\S}^{\M}(t').
\label{app:Markov}
\end{align}
Then, by assuming $\vRho_{\S}(t) = \vRho_{\S}^{\M}(t)$ for $\varDelta t \gg \tau_{\B}$, Eqs.~\eqref{app:nonMarkov} and \eqref{app:Markov} give 
\begin{align*}
&\oRho_{\S}^{\M}(t_0) = \oRho_{\S}(t_0) \\  
&- \textstyle {\int_{t_0}^{\infty} \dd t' \int_{t'-t_0}^{\infty} \dd \tau} 
 \dU_{\S}(t',t_0) \oC_{\S}(t',t'-\tau) \oU_{\S}(t',t_0) \oRho_{\S}(t_0),
\end{align*}
up to the second order in $\oH_{\S\B}$, where $t \to \infty$ is used since we have assumed $\varDelta t \gg \tau_{\B}$.
This equation means that in the Markovian QME we should use a `slipped' initial condition $\oRho_{\S}^{\M}(t_0) \equiv \SLP \oRho_{\S}(t_0)$
rather than the original one $\oRho_{\S}(t_0)$.
In other words, the Markovian QME is fully justified in combination with the slipped initial condition.
Then, transforming the integration $\int_{t_0}^{\infty} \dd t' \int_{t'-t_0}^{\infty} \dd \tau = \int_0^{\infty} \dd \tau \int_{t_0}^{\tau+t_0} \dd t'$, we find
\begin{widetext}
\begin{align}
\oRho_{\S}^{\M}(t_0) = \SLP \oRho_{\S}(t_0) 
= \left\{ 1 - \textstyle {\int_0^{\infty} \dd \tau \int_{t_0}^{\tau+t_0} \dd t'} \dU_{\S}(t',t_0) \oC_{\S}(t',t'-\tau) \oU_{\S}(t',t_0) \right\} \oRho_{\S}(t_0).
\label{app:SLP1}
\end{align}
Finally, by inserting Eq.~\eqref{app:temp1} into Eq.~\eqref{app:SLP1}, the specific form of the slippage superoperator is given by
\begin{align}
\SLP \oRho_{\S}(t_0) = &\oRho_{\S}(t_0) 
 + \textstyle {\sum_{\alpha, \beta} \int_0^{\infty} \dd \tau \int_{t_0}^{\tau+t_0} \dd t'} \left[ C_{\alpha\beta}(\tau)
 \{ \dvS_{\alpha}(t')\vS_{\beta}(t'-\tau) \oRho_{\S}(t_0) - \vS_{\beta}(t'-\tau)  \oRho_{\S}(t_0) \dvS_{\alpha}(t') \} + \Hc \right].
\label{app:SLP2}
\end{align}
Here, it is possible to apply the IEA $\vS_{\beta}(t'-\tau) \simeq \sum_{\epsilon(t')} e^{\ii\epsilon(t')\tau} \vS_{\beta}(\epsilon(t'); t')$
 [Eq.~\eqref{eq:IEA}] when $\tau_{\B} \ll \tau_{\A}(t')$.
However, we note that, even if $\tau_{\B} \ll \tau_{\A}(t')$ is not satisfied, there is no difficulty to numerically evaluate Eq.~\eqref{app:SLP2} because the integrand becomes negligible for $\tau \gg \tau_{\B}$ due to the decay of $C_{\alpha\beta}(\tau)$.
We also remark that Eq.~\eqref{app:SLP2} recovers the results by Gaspard and Nagaoka~\cite{Gaspard99} when the system Hamiltonian is assumed to be time-independent.

\section{Secular Approximation \label{sec:SA}} 
In this Appendix~\ref{sec:SA}, we discuss the SA to obtain the dissipator $\vD^{\I\E\S\A}$ [Eq.~\eqref{eq:IP::DIESA}] from $\vD^{\I\E\A}$ [Eq.~\eqref{eq:IP::DIEA}].
As we have seen in the time-independent case [Eqs.~\eqref{eq:std::D} and \eqref{eq:std::DSA}], to perform the SA, we have to extract the oscillating behaviors from $\dvS_{\alpha}(\epsilon(t); t)$ and $\vS_{\beta}(\epsilon(t); t)$ and have to average out the rapid oscillating terms in Eq.~\eqref{eq:IP::DIEA}.
For this purpose, we formally integrate $\frac{\dd}{\dd s} \vRho_{\S}(s) = \vD^{\I\E\A}_s \vRho_{\S}(s)$ over $s$ from $t$ to $t + \tau_{\X}(t)$,
\begin{multline}
\vRho_{\S}(t+\tau_{\X}(t)) -\vRho_{\S}(t) = 
 {\textstyle \sum_{\alpha,\beta} \int_0^{\tau_{\X}(t)} \dd s' \sum_{\epsilon(s'+t), \epsilon'(s'+t)} } \varGamma_{\alpha\beta}(\epsilon(s'+t)) \\
\times \{ \vS_{\beta}(\epsilon(s'+t); s'+t) \vRho_{\S}(s'+t) \dvS_{\alpha}(\epsilon'(s'+t); s'+t)
- \dvS_{\alpha}(\epsilon'(s'+t); s'+t) \vS_{\beta}(\epsilon(s+t); s'+t) \vRho_{\S}(s'+t) \} + \Hc,
\label{app:QME_SAtemp1}
\end{multline}
where $\tau_{\X}(t) \ge 0$ is a certain timescale at time $t$ and the integration variable has been changed from $s$ to $s' = t + s$ in the right-hand side.
Here, $\vS_{\beta}(\epsilon(s+t); s+t)$ can be rewritten as 
\begin{align}
\vS_{\beta}(\epsilon(s'+t); s'+t) = \du_{\S}(t, t_0)\du_{\S}(s'+t,t) \oS_{\beta}(\epsilon(s'+t)) \ou_{\S}(s'+t,t)\ou_{\S}(t,t_0),
\label{app:SA::EVseparation}
\end{align}
by using Eq.~\eqref{eq:EVprop1}.
In this equation, then, $\epsilon(s'+t) \simeq \epsilon(t)$ and $\ou_{\S}(s'+t,t) \simeq e^{-\ii\oH_{\S}(t)s'}$ are allowed if we assume 
\begin{align}
\tau_{\X}(t) \ll \tau_{\A}(t),
\label{app:SA::condition1}
\end{align}
due to Eqs.~\eqref{eq:AH1} and \eqref{eq:AH2}.
Equation~\eqref{app:SA::EVseparation} therefore results in
\begin{align}
\vS_{\beta}(\epsilon(s'+t); s'+t) &\simeq e^{-\ii\epsilon(t)s'}\vS_{\beta}(\epsilon(t); t),
\label{app:SA::Oscillate1}
\end{align}
and, similarly, we have
\begin{align}
\dvS_{\alpha}(\epsilon'(s'+t); s'+t) &\simeq e^{\ii\epsilon'(t)s'}\vS_{\beta}(\epsilon'(t); t).
\label{app:SA::Oscillate2}
\end{align}
These approximations have close analogies with the procedures for the IEA [Eqs.~\eqref{eq:EVseparation}--\eqref{eq:IEA}].
As a result, one can indeed find the oscillatory behaviors in Eqs.~\eqref{app:SA::Oscillate1} and \eqref{app:SA::Oscillate2}.
Inserting these equations into Eq.~\eqref{app:QME_SAtemp1}, we obtain
\begin{multline}
\vRho_{\S}(t+\tau_{\X}(t)) -\vRho_{\S}(t) \simeq 
 {\textstyle \sum_{\alpha,\beta}  \sum_{\epsilon(t), \epsilon'(t)} \varGamma_{\alpha\beta}(\epsilon(t)) \int_0^{\tau_{\X}(t)} \dd s'} 
e^{-\ii \{ \epsilon(t) - \epsilon'(t) \} s'}  \\
\times \{ \vS_{\beta}(\epsilon(t)) \vRho_{\S}(s'+t) \dvS_{\alpha}(\epsilon'(t))
- \dvS_{\alpha}(\epsilon'(t)) \vS_{\beta}(\epsilon(t)) \vRho_{\S}(s'+t) \} + \Hc.
\label{app:SA::Integrated2}
\end{multline}
\end{widetext}
There are two functions in the integrand that depend on the variable $s'$.
One is the density operator $\vRho(s'+t)$, the temporal change timescale of which is characterized by the relaxation timescale $\tau_{\R}(t)$ [Eq.~\eqref{eq:IEA::tauR}].
The other one is $e^{-\ii \{ \epsilon(t) - \epsilon'(t) \} s'}$ that corresponds to the beat between the two different Bohr frequencies $\epsilon(t)$ and $\epsilon'(t)$.
The intrinsic timescale of the oscillation period for $\epsilon(t) \ne \epsilon'(t)$ is described by $\tau_{\S}(t)$ [Eq.~\eqref{eq:IEA::tauS}] at time $t$.
Therefore, in the integral of Eq.~\eqref{app:SA::Integrated2}, $e^{-\ii \{ \epsilon(t) - \epsilon'(t) \} s'}$ for $\epsilon(t) \ne \epsilon'(t)$ oscillates rapidly when 
\begin{align}
\tau_{\S}(t) \ll \tau_{\X}(t),
\label{app:SA::condition2}
\end{align}
whereas $\vRho(s'+t)$ remains unchanged when 
\begin{align}
\tau_{\X}(t) \ll \tau_{\R}(t).
\label{app:SA::condition3}
\end{align}
In such a situation, indeed, the non-secular terms of $\epsilon(t) \neq \epsilon'(t)$ average out to zero.
As a result, only the secular terms, $\epsilon(t) = \epsilon'(t)$, survive in the integration of Eq.~\eqref{app:SA::Integrated2}.
Then, by retracing the steps [Eqs.~\eqref{app:QME_SAtemp1}-\eqref{app:SA::Integrated2}] in reverse order, we again arrive at Eq.~\eqref{app:QME_SAtemp1} but without non-secular terms.
This means that the dissipator $\vD^{\I\E\A}_t$ [Eq.~\eqref{eq:IP::DIEA}] is well approximated by $\vD^{\I\E\S\A}_t$, where
\begin{multline*}
\vD^{\I\E\S\A}_t\oRho_{\S} \equiv \sum_{\epsilon(t)} \sum_{\alpha, \beta} \varGamma_{\alpha\beta}(\epsilon(t)) 
 \{ \vS_{\beta}(\epsilon(t); t) \vRho_{\S} \dvS_{\alpha}(\epsilon(t); t) \\
- \dvS_{\alpha}(\epsilon(t); t)  \vS_{\beta}(\epsilon(t); t) \vRho_{\S}\} + \Hc.
\end{multline*}
This equation is nothing but Eq.~\eqref{eq:IP::DIESA}.
As is obvious from the above discussion, the SA is justified when the three conditions, $\tau_{\X}(t) \ll \tau_{\A}(t)$ [Eq.~\eqref{app:SA::condition1}], $\tau_{\S}(t) \ll \tau_{\X}(t)$ [Eq.~\eqref{app:SA::condition2}], and $\tau_{\X}(t) \ll \tau_{\R}(t)$ [Eq.~\eqref{app:SA::condition3}], are all satisfied.
Therefore, we need
\begin{align*}
\tau_{\S}(t) \ll \tau_{\R}(t) \quad \text{and} \quad \tau_{\S}(t) \ll \tau_{\A}(t),
\end{align*}
simultaneously, to perform the SA.
Thus, Eq.~\eqref{eq:IEA::SA} is obtained.


%

\end{document}